\documentclass[fleqn,usenatbib]{mnras}

\usepackage[T1]{fontenc}

\DeclareRobustCommand{\VAN}[3]{#2}
\let\VANthebibliography\thebibliography
\def\thebibliography{\DeclareRobustCommand{\VAN}[3]{##3}\VANthebibliography}

\newcommand{\ks}{km~s$^{-1}$}
\newcommand{\ms}{M$_{\odot}$}
\newcommand{\rs}{R$_{\odot}$}
\newcommand{\oc}{$O\!-\!C$}

\newcommand{\ond}{Ond\v{r}ejov}

\newcommand{\valmez}{Vala\v{s}sk\'e Mezi\v{r}\'{\i}\v{c}\'{\i}}
\newcommand{\vev}{Veversk\'a B\'{\i}t\'y\v{s}ka}
\newcommand{\n}{N745}

\usepackage{graphicx}	% Including figure files
\usepackage{amsmath}	% Advanced maths commands
\usepackage{amssymb}	% Extra maths symbols
\usepackage{newtxtext,newtxmath}
\title[A photometric study of NSVS~7453183]
{A photometric study of NSVS~7453183: 
a probable quadruple system \\ with long-term surface activity }

\author[L. \v{S}melcer et al.]{
L. \v{S}melcer,$^{1,7}$
M. Wolf,$^2$\thanks{E-mail: marek.wolf@mff.cuni.cz}
H. Ku\v{c}\'akov\'a,$^{2,3,4,7}$
P. Zasche,$^2$
J. K\'ara,$^2$
K. Hornoch,$^3$
M. Zejda,$^5$ 
and R.F. Auer$^{6,7}$
\\
$^{1} $\valmez\ Observatory, Vset\'{\i}nsk\'a 78, CZ-757~01 \valmez, Czech Republic \\
$^{2 }$Astronomical Institute, Faculty of Mathematics and Physics, Charles University Prague, 
V Hole\v{s}ovi\v{c}k{\'a}ch 2, CZ-180 00 Praha 8, Czech Republic\\
$^{3} $Astronomical Institute, Academy of Sciences, Fri\v{c}ova 298,  CZ-251~65 \ond, Czech Republic \\
$^{4}$Research Centre for Theoretical Physics and Astrophysics, Institute of Physics, Silesian University in Opava, Bezru\v{c}ovo n\'am. 13, CZ-746 01 Opava, Czech Republic \\
$^{5} $Department of Theoretical Physics and Astrophysics, Masaryk University, Kotl\'{a}\v{r}sk\'{a} 2, CZ-611~37~Brno, Czech Republic \\
$^{6} $South-Moravian Observatory, Chud\v cice 273, CZ-64771~\vev, Czech Republic \\
$^{7} $Czech Astronomical Society, Variable Star and Exoplanet Section, V\'ide\v{n}sk\'a~1056, CZ-142~00~Praha~4, Czech Republic
}

\date{Accepted 2023 January 3. Received 2022 December 6; in original form 2022 October 11}

\pubyear{2023}

\begin{document}
\label{firstpage}
\pagerange{\pageref{firstpage}--\pageref{lastpage}}
\maketitle

\begin{abstract}
The $VRC$ light curves were regularly measured for the eclipsing binary NSVS~7453183 as a part of our long-term observational project for studying of low-mass eclipsing binaries with a short orbital period and surface activity.
The {\sc Tess} light curve solution in {\sc Phoebe} results to the detached configuration, where the temperature of primary component was adopted  
to $T_1$ = 4300 K according to the SED approximation. 
It gives us $T_2 =$ 4080 $\pm$ 100 K for the secondary component. The spectral type of the primary component was estimated to be K6 and the photometric mass ratio was derived $q = 0.86$. 
We confirm presence of the third body in this system, a stellar companion with a minimal mass 0.33 \ms\ orbiting the eclipsing pair with a short period about 425 days, and propose the next, fourth body with a longer orbiting period of about 12 years, probably a brown dwarf with the minimal mass of 50 M$_{\rm Jup}$.
The hierarchical structure ((1+1)+1)+1 of this quadruple system is assumed. 
Characteristics and temporal variations of the dark region on the surface of 
the primary component were estimated. The average migration speed of about 10 deg/month
was found during years 2020-2022.
 
%The properties of binary components are compared with previously known similar systems  \dots
\end{abstract}

\begin{keywords}
binaries: close -- 
binaries: eclipsing -- 
binaries: low-mass -- 
stars: activity --
stars: fundamental parameters -- 
stars: individual: NSVS~7453183 
\end{keywords}

%%%%%%%%%%%%%%%%%   1. Intro  %%%%%%%%%%%%%%%%%%%%%%%%%%%%%%%%%
\section{Introduction}
\label{sec:intro}

The most common and frequent stars in our Galaxy are late-type and low-mass stars,
dwarfs with masses below 1.0 \ms.
Moreover, current observations of low-mass stars show a discrepancy
between estimated and modeled parameters, where the models
give 5–10 \% smaller radii and higher temperatures than observations \citep{2000ARA&A..38..337C, 2010ASPC..435..141M,2015ApJ...804...64M}.
Many eclipsing binaries (EB) display a periodic variation of their mid-eclipse times 
caused by various phenomena.
One is the light-time effect (LITE) associated with the presence of a third body orbiting the eclipsing pair \citep{1952ApJ...116..211I, 1959AJ.....64..149I}, the second alternative is magnetically-induced gravitational modulation caused by an active star in the system \citep{1992ApJ...385..621A}. The fundamental properties of late-type stars in EB were recently summarized by \cite{2022Galax..10...98M}, who confirmed the above mentioned
discrepancy and emphasize importance of low-mass stars to precision o our stellar evolution models.

Like our Sun, the low-mass stars are also affected by chromospheric activity caused 
by a strong magnetic field. This variable activity has been
frequently observed as flares, dark or bright spots on the surface 
and plays important role for determination of precise physical parameters, 
esp. their radii and temperatures.
For the first time, the spot activity in stars was announced by \cite{1950AJ.....55...69K} who detected the sinusoid-like variations in the light curve of eclipsing binary 
YY~Gem .

%Long-term photometric monitoring of low-mass eclipsing binaries is very useful tool for studying %star spot parameters (their surface, coordinates, sizes and temperatures), their evolution and %statistical properties. 

%%%%%%%%%%%%%%%%%%%%%%%%%%%   N745   %%%%%%%%%%%%%%%%%%%%%%%%%%%%%%%%%%%%%%%%%
%%%%%%%%%%%%%%%%%%%%%%%%%%%%%%%%%%%%%%%%%%%%%%%%%%%%%%%%%%%%%%%%%%%%%%%%%%%%%%

The low-mass eclipsing binary NSVS~07453183 (in {\sc Simbad} also as NSVS~07446320, 2MASS~J09161228+3615335, hereafter \n) is rather faint northern object with a short orbital period of about 9 hours. Its variability was discovered by \cite{2005IAPPP.101...38M} using the publicly available 
{\it Northern Sky Variability Survey} (NSVS, Wozniak et al. 2004). 
First photometric solution of \n\ was presented by \cite{2007JSARA...1....7C}
who derived preliminary masses and dimensions of components ($M_1 = 0.68, M_2 = 0.73$ \ms, $R_1 = 0.72 , R_2 = 0.79$ \rs). Using the program Eclipsing Light Curves (ELC)
they also found two spots on the surface of the hotter and more massive component with longitudes -33 and +17 deg and angular radii 100 and 9.3 deg, respectively.
Their analysis supports the current findings that low-mass stars have greater 
radii than models predict, most likely due to the presence of strong magnetic fields.
Next BVR photometry and period analysis of \n\  were presented by \cite{2014MNRAS.442.2620Z}, who also detected a flare-like event. 
Later, the period analyses of \n\ was presented by \cite{2016A&A...587A..82W}. They found a probable third body in this system, a stellar companion with a minimal mass 0.41 \ms\ orbiting the eclipsing pair with a short period of about 418 days. The following linear ephemeris were proposed in that paper for the current use:
\medskip

Pri.Min. = HJD 24 53456.85994 + $0\fd36696751 \cdot E$.
\medskip

\noindent
{\sc Gaia} DR2 astrometric and phometric data on \n\ are summarized in Table~\ref{tg} 
\citep{2020arXiv201201533G}. The distance to the system was derived to be $d = 314$ pc.
 To our knowledge, no modern photometric or spectroscopic analyses of this short-period eclipsing binary exists so far.

\begin{table}
\begin{center}
\caption{{\sc Gaia} DR2 astrometric and photometric data on \n.}
\label{tg}
\begin{tabular}{cc}
\hline\hline\
 Parameter              &  Value     \\
\hline
$\alpha_{2000}$ [h m s] & 09 16 12.27 \\
$\delta_{2000}$ [d m s] & +36 15 33.5 \\
pm $\alpha$ [mas/yr]    & --28.52 $\pm$ 0.07 \\
pm $\delta$ [mas/yr]    & --11.75 $\pm$ 0.06 \\
parallax $\pi$ [mas]    & 3.311   $\pm$ 0.074  \\
\hline
$V$ [mag]                 & 13.012  $\pm$ 0.038 \\
$G$ [mag]                 & 13.000  $\pm$ 0.0074 \\
$BP$ [mag]                & 13.629  $\pm$ 0.0258\\
$RP$ [mag]                & 12.287  $\pm$ 0.0207  \\
$J$ [mag]                 & 11.300  $\pm$ 0.022 \\
$H$ [mag]                 & 10.740  $\pm$ 0.016 \\
$K$ [mag]                 & 10.608  $\pm$ 0.016 \\
\hline
\end{tabular}
\end{center}
\end{table}

The paper is organized as follows: 
in Section~\ref{sec:obs} we present the observational data used in the analysis. 
The orbital period study and the light-time effect are described in Section~\ref{sec:period}. The photometric solution is given in Section~\ref{sec:curve} and 
discussion of the flare activity of \n\  is given in Section~\ref{sec:flare}. 
The paper is finished by discussion (Section~\ref{sec:discus}) and our conclusions 
are summarized in Section~\ref{sec:concl}.

%%%%%%%%%%%%%%%%%%%%%%%% 2. Observations  %%%%%%%%%%%%%%%%%%%%%%%%%%%%%%
\section{Observations}
\label{sec:obs}
%%%%%%%%%%%%%%%%%%%%%%%%%%%%%%%%%%%%%%%%%%%%%%%%%%%%%%%%%%%%%%%%%%%%%%%%

In this section, we describe the properties and reduction process
of our observations of \n. We present our long-term ground-based photometry (\ref{sec:ground}), 
{\sc Tess} photometry (\ref{sec:tess}),  
and an older optical spectrum from Keck observatory (\ref{sec:Keck}).

\begin{figure}
\begin{center}
\includegraphics[width=0.85\linewidth]{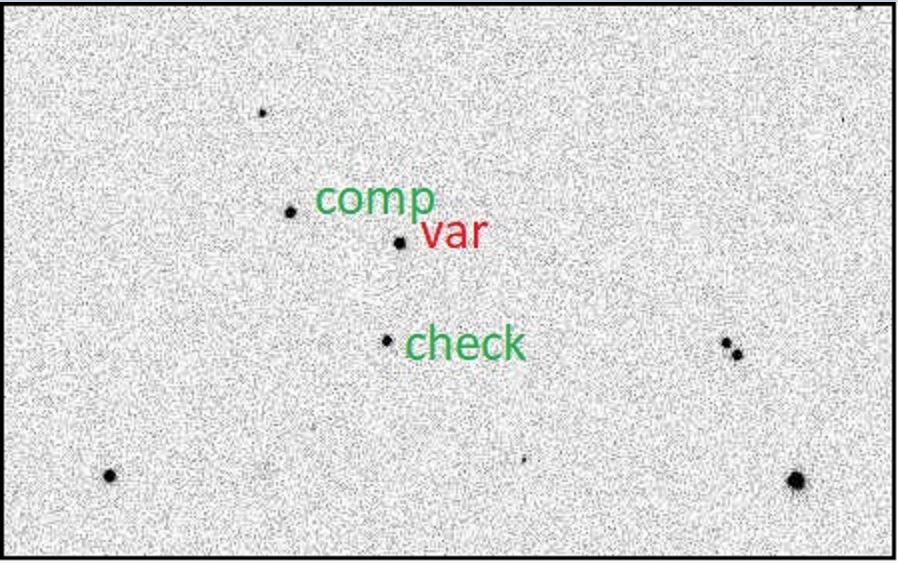}
\caption[ ]{The finding chart of \n, the dimensions are approx. $8 \times 5$ arcmin, north is up. The position of variable, comparison and check stars are denoted.} 
\label{fc}
\end{center}
\end{figure}

\begin{table}
\begin{center}
\caption{Coordinates of selected stars in the field of \n, see Figure~\ref{fc}. }
\label{t0}
\begin{tabular}{ccccc}
\hline\hline
Object & $\alpha_{2000}$ [h:m:s] & $\delta_{2000}$ [d:m:s] & $V$ [mag] & USNO-B1.0 \\
\hline
N745  (var)  & 09 16 12.27  &  +36 15 33.51 &  13.0  & 1262-0172833 \\
comparison   & 09 16 17.00  &  +36 15 49.41 &  13.25 & 1262-0172847	\\
check        & 09 16 12.74  &  +36 14 42.27 &  13.91 & 1262-0172837 \\
\hline
\end{tabular}
\end{center}
\end{table} 

\subsection{Ground-based photometry}
\label{sec:ground}

Since 2009 the time-resolved CCD photometry of \n\ mostly during eclipses has been regularly obtained at several observatories in Czech Republic. 
The focused photometric campaign has been
initiated in April 2019 to determine the eruption activity and flare frequency and to derive their characterization and relatively large amount of photometric data was obtained.

\begin{itemize}
 
\item Most observations in our campaign were obtained at \valmez\ observatory, Czech Republic, 
in 2014, and more systematically from January 2019 to November 2022. 
Three different telescopes, CCD camera a filters were used regularly each clear night:
Celestron SCT 280/1765, camera MII G2-4000 with filter B,
Sky-Watcher NWT 254/1200, camera CCD QHY 174 and filter C (clear), and
Celestron SCT 355/2460, camera MII G2-1600 with filter V.

\item Since 2009 semi-regular observations were performed at \ond\ observatory, Czech Republic, where the Mayer 0.65-m ($f/3.6$) reflecting telescope with the CCD camera MII G2-3200 and $VR$ photometric filters 
were used.

\item Important additional photometry was obtained at Masaryk University Observatory in Brno, Czech Republic, during three separate epochs: in March/April 2010, February/March 2020 and April/May 2022. The 0.60-m reflecting telescope with the CCD camera MII G4-16000 and VR photometric filters were included.

\item Supplementary photometry was obtained at South-Moravian Observatory of R.F.A. in \vev, Czech Republic, where 0.2-m ($f/4$) Newtonian telescope with the CCD camera MII G2-1600 and C (clear) filter were used during March/April 2021.

\end{itemize}

\noindent
The CCD observations were reduced in a standard way. 
The {\sc C-Munipack}\footnote{Package of software utilities 
for reducing astronomy CCD images, current version 2.1.31, 
available at \url{http://c-munipack.sourceforge.net/}},
a synthetic aperture photometry software, was routinely used for our time-series photometry.
At \ond\ the {\sc Aphot} \citep{1994ExA.....5..375P}, synthetic aperture photometry and astrometry software were used for time series.
Time-series were constructed by computing the magnitude difference between the variable and a nearby comparison and check stars, see Fig.~\ref{fc} and Table~\ref{t0},
the heliocentric correction was applied. 
The uncertainties of photometric measurements at smaller telescopes were always about 0.01 -- 0.02~mag.
Computers at the 65-cm telescope are synchronized using a time-server provided by  \url{http://ntp.cesnet.cz}  every two minutes. These corrections are usually in order of $10^{-3}$ seconds.
Together at all observatories it was obtained nearly 60~000 frames in V, R, I and C filters.

%%%%%%%%%%%%%%%%%%%%%%%%%%%%%%%%%%
\subsection{TESS photometry}
\label{sec:tess}
%%%%%%%%%%%%%%%%%%%%%%%%%%%%%%%%%

As a northern object \n\ was measured twice by the {\it Transiting Exoplanet Survey Satellite} ({\sc Tess}, \cite{2015JATIS...1a4003R})  in Sector~21 (Jan/Feb 2020) and Sector~48 (Jan/Feb 2022). 
We used the MAST~\footnote{MAST: Barbara A. Mikulski Archive for Space Telescopes, \\ \url{https://mast.stsci.edu/portal/Mashup/Clients/Mast/Portal.html}} database to download the photometric time-series. 
These high-quality light curves allowed us the precise mid-eclipse time determination as well as  modeling of the system.

The new times of primary and secondary minima and their uncertainties were generally determined by fitting the light curve by Gaussians or polynomials of the third or fourth order; we used the least-squares method. They are listed in Table~\ref{t1} and \ref{tess}, where epochs were computed according to the following improved linear ephemeris:

\begin{center}
Pri. Min. = BJD 24 53456.8597 + 0\fd366967684 $\cdot E$. \\
\end{center}

\noindent
Because the {\sc Tess} data are provided in the Barycentric Julian Date Dynamical Time 
(BJD$_{\rm TDB}$), all our times in Table~\ref{t1} were first transformed to this time 
scale using the often used Time Utilities of Ohio State University\footnote{\url{http://astroutils.astronomy.ohio-state.edu/time/} } 
\citep{2010PASP..122..935E}.

\begin{figure}
\centering
   \includegraphics[width=0.85\linewidth]{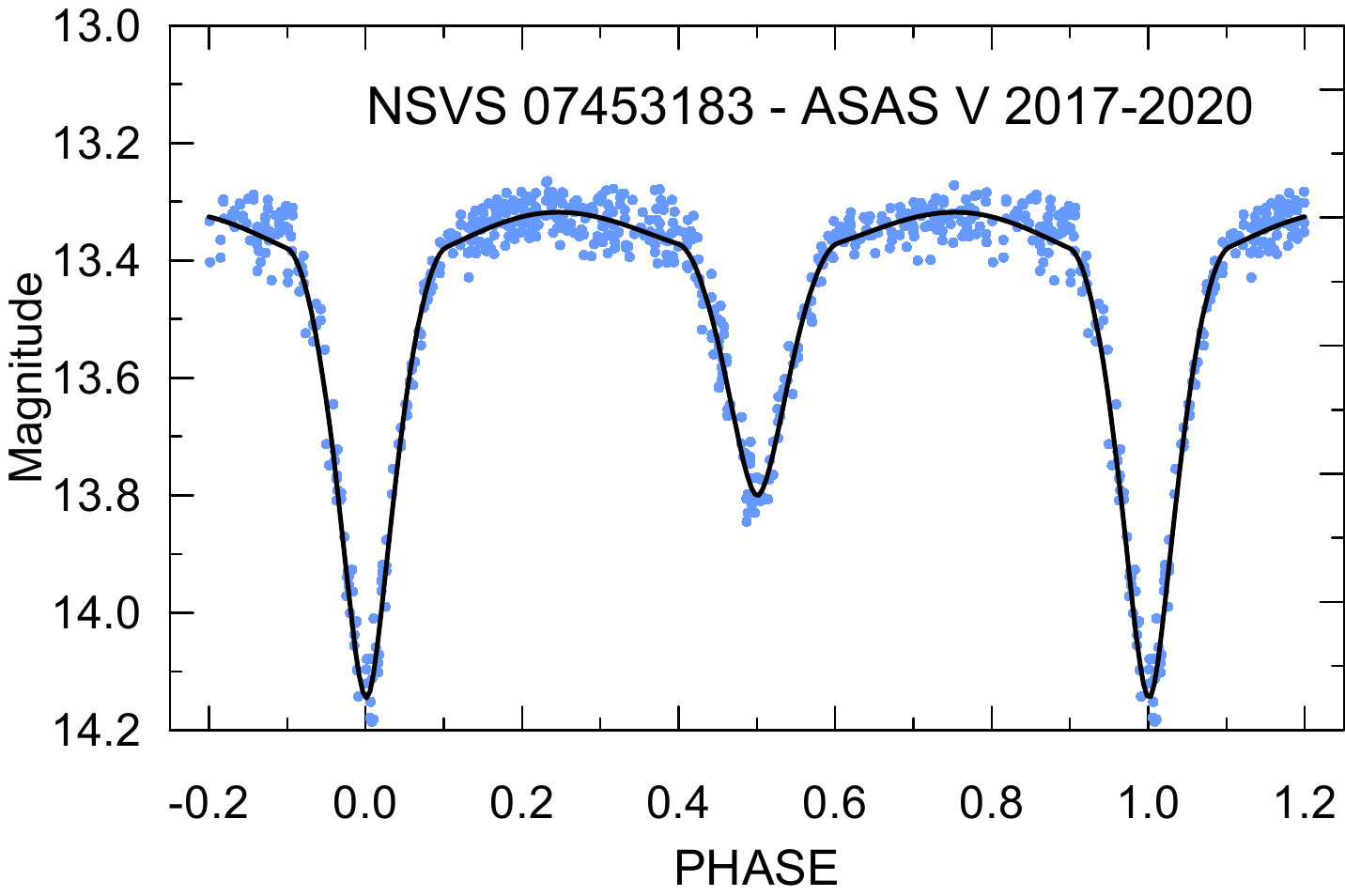} \\
    \caption{For illustration, the mean ASAS SN-V light curve of \n\ obtained 
    during 2017-2020 and its solution in {\sc Phoebe}. Due to relatively large
    scatter of data no spots on the surface of both components were employed.}
    \label{ASASlc}
\end{figure}

\n\ is also included in the ASAS-SN database \citep{2014ApJ...788...48S} as an object ASASSN-V J091612.17 +361533.7. Unfortunately, the large scatter of these data does not allow us to determine the reliable solution of the light curve nor precise parameters of possible surface spots. The ASASSN-V light curve of \n\ obtained during 2017-2020 is plotted in Fig.~\ref{ASASlc}.

%%%%%%%%%%%%%%%%%%%%%%%%%%%%%%%%%%%%%%%%%
\subsection{Keck spectrum}
\label{sec:Keck}
%%%%%%%%%%%%%%%%%%%%%%%%%%%%%%%%%%%%%%%%%

\begin{figure}
\begin{center}
\includegraphics[width=0.95\linewidth]{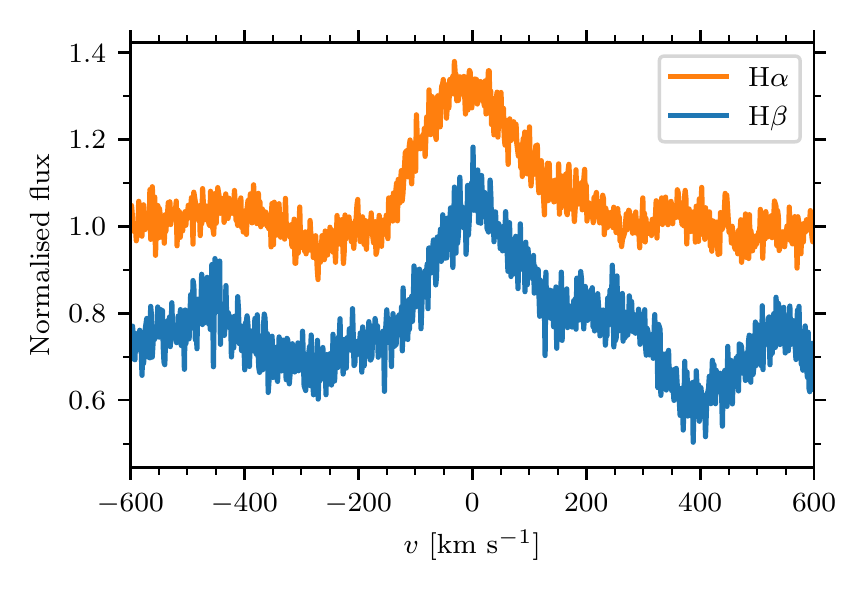}
\caption[ ]{The broad hydrogen H$\alpha$ and H$\beta$ line profiles of \n\  obtained at Keck Observatory in January 2012.} 
\label{keckspec}
\end{center}
\end{figure}

One spectrum of \n\ obtained on January 1, 2012, is immediately available in the Keck Observatory Archive (KOA)~\footnote{KOA Data Access Service - v18.5, \url{https://koa.ipac.caltech.edu/cgi-bin/KOA/nph-KOAlogin}}.
The HIRES spectrograph with 180 sec exposure time was used. The whole calibrated spectrum covers the wavelengths 4450 -- 8910 \AA\ and clearly shows the emission lines of the Balmer series of hydrogen (see Fig.~\ref{keckspec} for illustration).

\begin{table*}
\caption{New ground-based times of primary and secondary eclipses of \n\ obtained since January 2019.}
\label{t1}
\begin{tabular}{llclcllclcllclc}
\hline\hline
BJD & Error & Filters & Epoch & O & BJD & Error & Filters & Epoch & O & BJD & Error & Fil. & Epoch & O \\
-24 00000   & [day]   &   &  &    & -24 00000   & [day] &  &  &  & -24 00000 & [day] \\      
\hline
58503.39892   & 0.00005 & R     & 13752.0    & 2 &     59269.44233   & 0.0003 & C,V,R   & 15839.5 &  1 &      59595.49524   & 0.0001 & C &    16728.0 &  1 \\  
58506.3332    & 0.0003 & C      & 13760.0    & 1 &     59269.62762   & 0.0002 & C,V,R   & 15840.0 &  1 &      59599.34786   & 0.0002 & C &    16738.5 &  1 \\  
58590.3722    & 0.0001 & C      & 13989.0    & 1 &     59271.64319   & 0.0001 & V,R       & 15845.5 & 2 &     59624.30126   & 0.0001 & C,R &  16806.5 &  1 \\  
58591.4732    & 0.0001 & C      & 13992.0    & 1 &     59275.31363   & 0.0003 & C,B,V,R,I & 15855.5 &  1 &    59624.48479   & 0.0001 & C,R &  16807.0 &  1 \\  
58888.53286  & 0.0002 & V,R  & 14801.5 &  4 & 59276.41449   & 0.0002 & C,B,V,R,I & 15858.5 &  1 &    59625.40217   & 0.0001 & C &    16809.5 &  1 \\  
58926.51274 & 0.0002 & V,R  & 14905.0 & 4 & 59276.59977   & 0.0003 & C,B,V,R,I & 15859.0 &  1 &    59625.58609   & 0.0001 & C &    16810.0 &  1 \\  
58935.31922   & 0.0001 & R      & 14929.0    & 2 &     59281.37048   & 0.0001 & C,V,R   & 15872.0 &  1 &      59628.33789   & 0.0001 & C &    16817.5 &  1 \\  
58947.43041   & 0.0001 & C      & 14962.0    & 1 &     59281.55226   & 0.0001 & C,V,R   & 15872.5 &  1 &      59639.53054   & 0.0001 & C &    16848.0 &  1 \\  
58948.34981   & 0.0001 & C      & 14964.5    & 1 &     59284.30628   & 0.0001 & C,V,R   & 15880.0 &  1 &      59640.26458   & 0.0001 & C &    16850.0 &  1 \\  
58951.53129   & 0.0001 & C,V    &  14973.0   & 1 &     59284.48818   & 0.0002 & C,V,R   & 15880.5 &  1 &      59640.44734   & 0.0001 & C &    16850.5 &  1 \\  
58952.38635   & 0.0001 & C,B,V  &  14975.5  &  1 &     59286.32282   & 0.0001 & C,V,R   & 15885.5 &  1 &      59641.36568   & 0.0001 & C &    16853.0 &  1 \\  
58955.32222   & 0.0002 & C,B,V  &  14983.5  &  1 &     59286.50824   & 0.0001 & C       & 15886.0 &  1 &      59641.54840   & 0.0001 & C &    16853.5 &  1 \\  
58955.50421   & 0.0002 & C,B,V  &  14984.0  &  1 &     59294.39634   & 0.0002 & C,V,R   & 15907.5 &  1 &      59647.41969   & 0.0001 & C &    16869.5 &  1 \\  
58956.42338   & 0.0001 & C,B,V  &  14986.5  &  1 &     59294.58117   & 0.0001 & C,R &  15908.0 &  1 &         59648.33769   & 0.0001 & C &    16872.0 &  1 \\  
58959.35896   & 0.0002 & C,B,V  &  14994.5  &  1 &     59299.35240   & 0.0001 & C &    15921.0 &  1 &         59649.43873   & 0.0001 & C &    16875.0 &  1 \\  
58960.45965   & 0.0002 & C,B,V  &  14997.5  &  1 &     59300.26863   & 0.0001 & C &    15923.5 &  1 &         59649.62123   & 0.0001 & C &    16875.5 &  1 \\  
58961.37575   & 0.0001 & B,V    & 15000.0 &    1 &     59300.45318   & 0.0001 & C &    15924.0 &  1 &         59650.35565   & 0.0001 & C &    16877.5 &  1 \\  
58962.47694   & 0.0002 & B,V    & 15003.0 &    1 &     59304.30487   & 0.0001 & C &    15924.5 &  1 &         59650.53970   & 0.0001 & C &    16878.0 &  1 \\  
58963.39574   & 0.0002 & C,B,V  &  15005.5 &   1 &     59304.48998   & 0.0001 & C &    15935.0 &  1 &         59651.27353   & 0.0001 & C &    16880.0 &  1 \\  
58966.33150   & 0.0002 & B,V    & 15013.5 &    1 &     59304.30386   & 0.0003 & C &    15934.5 &  3 &         59651.45660   & 0.0001 & C &    16880.5 &  1 \\  
58967.43226   & 0.0002 & C,B,V  &  15016.5 &   1 &     59304.48892   & 0.0003 & C &    15935.0 &  3 &         59651.64037   & 0.0001 & C &    16881.0 &  1 \\  
58968.34814   & 0.0001 & C,B,V  &  15019.0 &   1 &     59305.40558   & 0.0001 & C &    15937.5 &  1 &         59652.37491   & 0.0001 & C &    16883.0 &  1 \\  
58990.36657   & 0.0001 & C,V    & 15079.0 &    1 &     59309.44237   & 0.0001 & C &    15948.5 &  1 &         59652.55782   & 0.0002 & C &    16883.5 &  1 \\  
59127.61190   & 0.0001 & V,R,I &   15453.0 &   1 &     59311.27770   & 0.0001 & C &    15953.5 &  1 &         59657.32840   & 0.0001 & C &    16896.5 &  1 \\  
59161.55480   & 0.0002 & V,R,I &   15545.5 &   1 &     59311.46251   & 0.0001 & C &    15954.0 &  1 &         59657.51181   & 0.0001 & C &    16897.0 &  1 \\  
59172.56298   & 0.0003 & B,V,R,I & 15575.5 &   1 &     59313.29747   & 0.0001 & C &    15959.0 &  1 &         59659.34686   & 0.0001 & C &    16902.0 &  1 \\  
59174.58364   & 0.0001 & C &       15581.0 &   1 &     59313.47910  &  0.0001 & C &    15959.5 &  1 &         59659.52992   & 0.0001 & C &    16902.5 &  1 \\  
59175.49955   & 0.0001 & C &       15583.5 &   1 &     59313.29643  &  0.0003 & C &    15959.0 &  3 &         59660.44757   & 0.0001 & C &    16905.0 &  1 \\  
59178.43494   & 0.0003 & B,V,R,I & 15591.5 &   1 &     59328.34354  &  0.0001 & C &    16000.0 &  1 &         59661.36466   & 0.0001 & C &    16907.5 &  1 \\  
59178.62044   & 0.0002 & B,V,R,I & 15592.0 &   1 &     59330.35988  &  0.0001 & C &    16005.5 &  1 &         59661.54864   & 0.0001 & C &    16908.0 &  1 \\  
59184.49144   & 0.0001 & V,R     & 15608.0 &   1 &     59332.38020  &  0.0001 & C &    16011.0 &  1 &         59662.46558   & 0.0001 & C &    16910.5 &  1 \\  
59184.67316   & 0.0001 & V,R     & 15608.5 &   1 &     59333.29582  &  0.0001 & C &    16013.5 &  1 &         59663.38343   & 0.0001 & C &    16913.0 &  1 \\  
59185.40729   & 0.0003 & C,V,R   & 15610.5 &   1 &     59333.48106  &  0.0001 & C &    16014.0 &  1 &         59663.56644   & 0.0001 & C &    16913.5 &  1 \\  
59185.59294   & 0.0001 & C,V,R   & 15611.0 &   1 &     59338.43333  &  0.0001 & C &    16027.5 &  1 &         59666.31891   & 0.0001 & C &    16921.0 &  1 \\  
59192.56481   & 0.0001 & C,V,R   & 15630.0 &   1 &     59343.38965  &  0.0001 & C &    16041.0 &  1 &         59666.50212   & 0.0001 & C &    16921.5 &  1 \\  
59210.36019   & 0.0003 & V,R     & 15678.5 &   1 &     59344.30559  &  0.0002 & C &    16043.5 &  1 &         59681.36442   & 0.0001 & C &    16962.0 &  1 \\  
59210.54617   & 0.0001 & V,R     & 15679.0 &   1 &     59344.49016   & 0.0001 & C &    16044.0 &  1 &         59681.54797   & 0.0001 & C &    16962.5 &  1 \\  
59225.40604   & 0.0002 & C,V,R   & 15719.5 &   1 &     59345.40593   & 0.0001 & C &    16046.5 &  1 &         59682.28176   & 0.0002 & C &    16964.5 &  1 \\  
59225.59153   & 0.0001 & C       & 15720.0 &   1 &     59368.34349   & 0.0001 & C &    16109.0 &  1 &         59682.46553   & 0.0001 & C &    16965.0 &  1 \\  
59242.65322   & 0.0002 & C,V,R,I & 15766.5 &   1 &     59370.36039   & 0.0001 & C &    16114.5 &  1 &         59684.30031   & 0.0001 & C &    16970.0 &  1 \\  
59260.26792   & 0.0003 & B,V,R,I & 15814.5 &   1 &     59488.52565   & 0.0001 & C &    16436.5 &  1 &         59689.43762   & 0.0001 & C &    16984.0 &  1 \\  
59260.45347   & 0.0002 & B,V,R,I & 15815.0 &   1 &     59496.59883   & 0.0001 & C &    16458.5 &  1 &         59691.45616   & 0.0002 & C &    16989.5 &  1 \\  
59260.63513   & 0.0003 & B,V,R,I & 15815.5 &   1 &     59512.56291   & 0.0001 & C,R &  16502.0 &  1 &         59698.42869   & 0.0001 & C &    17008.5 &  1 \\  
59264.30463   & 0.0002 & C,B,V,R,I & 15825.5 & 1 &     59523.57177   & 0.0001 & C &    16532.0 &  1 &         59699.34586   & 0.0001 & C &    17011.0 &  1 \\  
59265.40487   & 0.0001 & V         & 15828.5 & 2 &     59543.56987   & 0.0001 & C &    16586.5 &  1 &         59710.35487   & 0.0001 & C &    17041.0 &  1 \\  
59265.40474   & 0.0001 & R         & 15828.5 & 2 &     59552.56098   & 0.0001 & C &    16611.0 &  1 &         59711.45596   & 0.0001 & C &    17044.0 &  1 \\  
59266.32367   & 0.0001 & V         & 15831.0 & 2 &     59586.32110   & 0.0001 & C &    16703.0 &  1 &         59712.37331   & 0.0001 & C,R &  17046.5 &  1 \\  
59266.32386   & 0.0001 & R         & 15831.0 & 2 &     59586.50400   & 0.0001 & C &    16703.5 &  1 &         59719.34562   & 0.0001 & C &    17065.5 &  1 \\  
59267.42574   & 0.0002 & C,B,V,R,I & 15834.0 & 1 &     59586.68804   & 0.0001 & C &    16704.0 &  1 &         59891.63920   & 0.0001 & R &    17535.0  & 1 \\ 
59267.60711   & 0.0002 & C,B,V,R,I & 15834.5 & 1 &     59587.42214   & 0.0001 & C &    16706.0 &  1 &         59896.59414   & 0.0002 & R &    17548.5  & 1 \\ 
59268.34107   & 0.0002 & C,V,R,I   & 15836.5 & 1 &     59587.60504   & 0.0001 & C &    16706.5 &  1 &         59913.47404   & 0.0002 & R &    17594.5  & 1 \\      
59269.26049   & 0.0002 & C,V,R    & 15839.0 &  1 &     59595.31112   & 0.0001 & C &    16727.5 &  1 &                       &        &   &              &   \\     
\hline
\end{tabular}
\medskip
Notes: O (observatory): 1 - \valmez, 2 - \ond, 3 - \vev, 4 - Brno
\end{table*}

\begin{table}
\caption{New times of primary and secondary eclipses of \n\ based on the {\sc Tess} photometry,
Sector 48, January 2022.}
\label{tess}
\begin{center}
\begin{tabular}{cc}
\hline\hline
BJD           &   Epoch     \\
-24 00000     &             \\    
\hline
59609.43960   &   16766.0   \\ 
59609.80643   &   16767.0   \\ 
59609.98971   &   16767.5   \\ 
59610.17350   &   16768.0   \\ 
59610.35645   &   16768.5   \\ 
59610.54048   &   16769.0   \\ 
59610.72336   &   16769.5   \\ 
59610.90737   &   16770.0   \\ 
59611.09031   &   16770.5   \\ 
59611.27434   &   16771.0   \\ 
59611.45730   &   16771.5   \\ 
59611.64139   &   16772.0   \\ 
59611.82448   &   16772.5   \\ 
59612.00834   &   16773.0   \\ 
\hline
\end{tabular}
\end{center}
\end{table}

%%%%%%%%%%%%%%%%%%%%%%%%%%%%%% 3. O-C diagram  %%%%%%%%%%%%%%%%%%%%%%%%%%%%%%
\section{Orbital period study}
\label{sec:period}
%%%%%%%%%%%%%%%%%%%%%%%%%%%%%%%%%%%%%%%%%%%%%%%%%%%%%%%%%%%%%%%%%%%%%%%%%%%%%

The period changes of \n\ were studied in detail last time in \cite{2016A&A...587A..82W}. 
To confirm our previous finding we continued in monitoring of eclipses to extend the \oc\ diagram and to reveal the nature of the rapid period changes. Based on the current data set extended by seven years of continuous measurements we propose a quadruple system with remarkable surface activity, where the eclipsing pair is orbiting by two additional bodies. Such a long-term project was possible only with the help of several advanced amateur observers.

%%%%%%%%%%%%%%%%%%%%%%%%%%%%%%
\subsection{Light-time effect}
%%%%%%%%%%%%%%%%%%%%%%%%%%%%%%

\begin{figure*}
\centering
\includegraphics[width=0.6\linewidth]{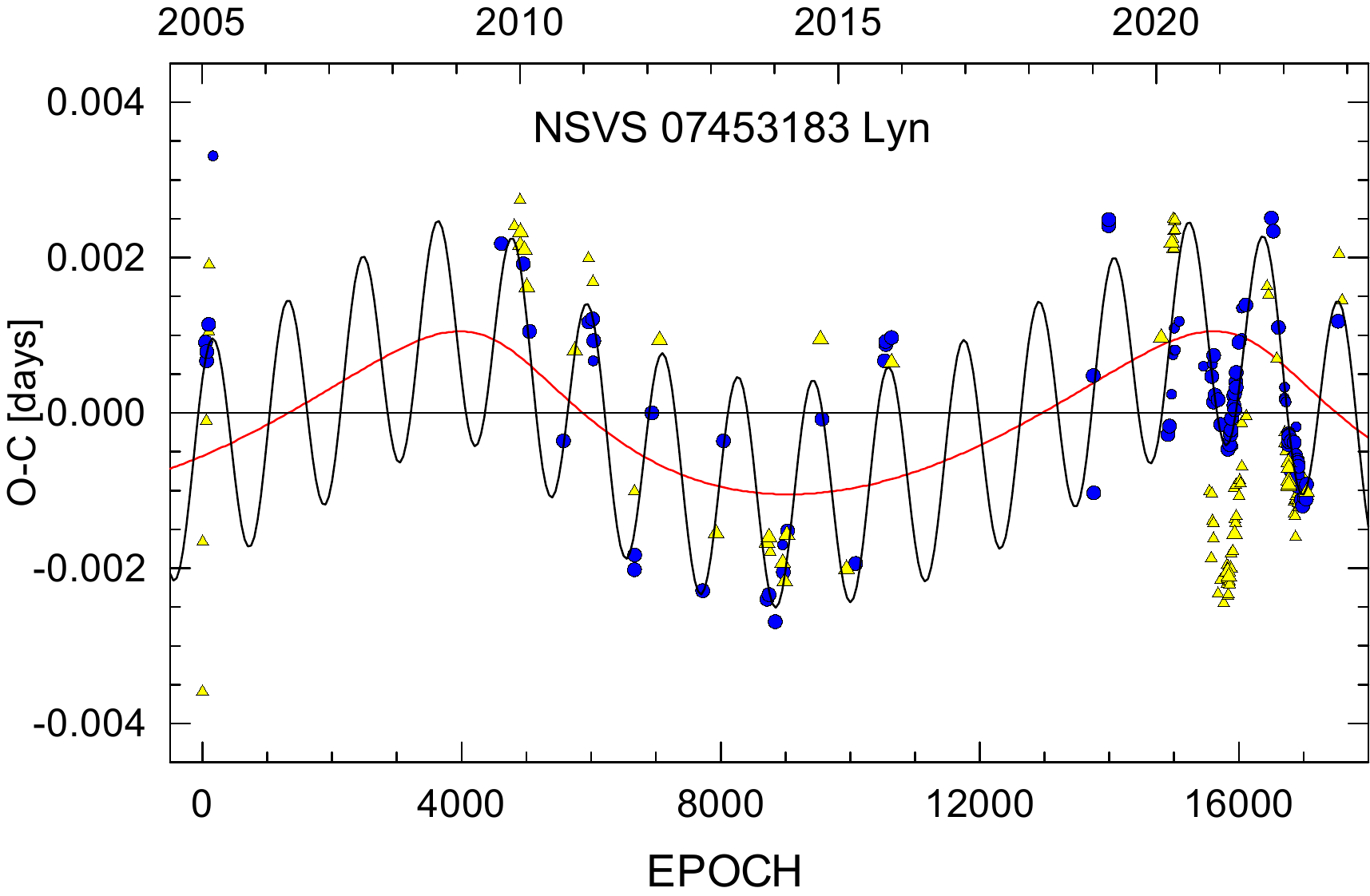}
\caption[ ]{The complete \oc\ diagram for the times of minimum of \n\ since its discovery in 2005.  The individual primary and secondary CCD minima obtained are denoted by blue circles and yellow triangles, respectively. Larger symbols correspond to the precise CCD measurements, which were used in our calculation of the LITE.
    The black sinusoidal curve represents the LITE with the short period of about 425~days. The red curve denotes long-term change with a period of 12 years caused probably by a fourth orbiting body. }
\label{n745oc}
\end{figure*}

\begin{figure}
\centering
\includegraphics[width=0.9\linewidth]{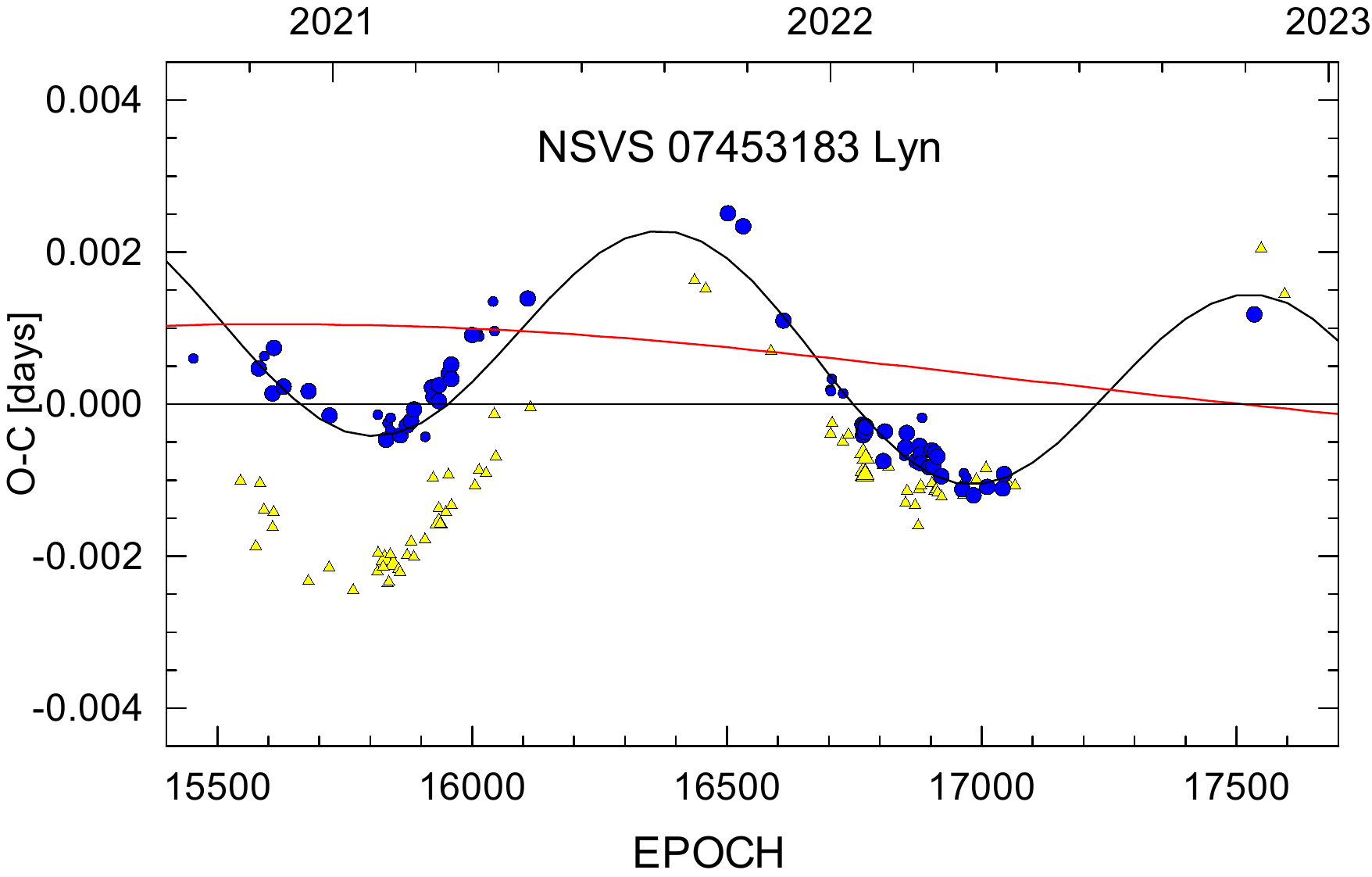}
\caption[ ]{The \oc\ diagram of \n\ in detail for minima obtained during two last seasons in 2021 and 2022. The black sinusoidal curve fitting the primary minima denotes the LITE with short period of 425 days, the red curve corresponds to LITE of the fourth body with period of 12 years. 
Note, that the secondary minima (yellow triangles) have a different course modulated by 
short-period LITE.  }
\label{n745ocd3}
\end{figure}

The light-time effect (LITE) in \n\ were studied traditionally
by means of an \oc\ diagram (or ETV curve) analysis.
As has been proven many times in the past, this simple and effective technique 
is a very powerful tool to investigate the multiplicity of stellar systems.
For a detailed description of LITE analyses see the original papers 
of \cite{1952ApJ...116..211I, 1959AJ.....64..149I, 1973A&AS...12....1F, 1990BAICz..41..231M, 2005ASPC..335.....S}.
The basic fitting equations are as follows. The observed deviations of mid-eclipse times  $(O-C)_{\rm obs}$ from the linear ephemeris is given by a superposition caused by 
a third and fourth bodies:

\medskip
$(O-C)_{\rm obs} = (O-C)_{\rm LITE,3} + (O-C)_{\rm LITE,4}.$

\medskip
\noindent
The light-travel time caused by a third body is given by

\begin{equation}
 (O - C)_3 = \frac{A_3}{\sqrt{1-e_3^2 \cos^2 \omega_3}}
      \biggl[ \frac{1-e_3^2}{1+e_3 \cos {\nu_3}} \sin({\nu_3} + \omega_3) + e_3 \sin \omega_3 \biggl] ,  
\end{equation}

\noindent where $e_3$ is the eccentricity of the third-body orbit,
$\omega_3$ the longitude of periastron and $\nu_3$ the true anomaly
of the third body. Similarly for the fourth body

\begin{equation}
 (O - C)_4 = \frac{A_4}{\sqrt{1-e_4^2 \cos^2 \omega_4}}
      \biggl[ \frac{1-e_4^2}{1+e_4 \cos {\nu_4}} \sin({\nu_4} + \omega_4) + e_4 \sin \omega_4 \biggl] ,  
\end{equation}

\noindent where $e_4$ is the eccentricity of the fourth-body orbit,
$\omega_4$ the longitude of periastron and $\nu_4$ the true anomaly
of the fourth body.
The observed semi-amplitudes $A_3$ and $A_4$ of the light-time curves (in days)
are

\begin{equation}
A_3 = \frac{a_{12,3}\sin i_3} {173.15} \, \sqrt{1-e_3^2 \cos^2 \omega_3}, 
\end{equation}

\begin{equation}
A_4 = \frac{a_{12,4}\sin i_4} {173.15} \, \sqrt{1-e_4^2 \cos^2 \omega_4},
\end{equation}

\noindent where $a_{12,3}$ and $a_{12,4}$ are the semi-major axes of the relative orbits of the eclipsing pair around the common center of mass
(in AU), $i_3$ and $i_4$ are inclinations of the third- and fourth body orbits. There are following independent variables to be determined in this procedure:

\medskip
\noindent
\hspace{1cm} $(A_3, P_3, T_3, e_3, \omega_3)$ for the LITE of the third body, and 

\smallskip
\noindent
\hspace{1cm} $(A_4, P_4, T_4, e_4, \omega_4)$ for the LITE of the fourth body,

\medskip 
\noindent
where $P_3$ and $P_4$ are orbital periods of the third and fourth bodies, $T_3$ and $T_4$ are times of periastron and $e_3$ and $e_4$ are eccentricities of their respective orbits. The linear ephemeris ($T_0, P_\mathrm{s}$) of the eclipsing pair are a part of this set of 12 unknown parameters. 

\begin{table}
\begin{center}
\caption{The improved LITE elements and the minimal masses of the possible third and fourth bodies.}
\label{t2}
\begin{tabular}{cccc}
\hline\hline
Parameter       & Unit  &    Value  \\
\hline
$T_0$           & BJD   & 24 53456.8597 (1)  \\
$P_s$           & day   & 0.366967681 (9)   \\
\hline
$A_3$           & days  &   0.00146 (5)    \\
$P_3$           & days  &   425.5 (7.8)   \\
$P_3$           & years &   1.165 (0.021)   \\
$e_3$           & --    &   0.0 (fixed)    \\
$\omega_3$      & deg   &   357.4 (1.2)  \\
$T_3$           &  JD   &  24~58930 (10)     \\
$f(M_3)$        & \ms   &   0.0120      \\
$M_{\rm 3, min}$ & \ms  &   0.33       \\
$K_3$           & \ks   &   6.5         \\
$A_{\rm dyn,3}$ &  days &   0.000 05    \\   
\hline
$A_4$           & days  &   0.00105 (15)    \\
$P_4$           & days  &   4271 (350)  \\
$P_4$           & years &   11.7 (0.9)    \\
$e_4$           &  --   &   0.40 (0.15)   \\
$\omega_4$      & deg   &   122.6 (2.5)        \\
$T_4$           & JD    &   24~55138 (30)   \\
$f(M_4)$        & \ms   &   0.000 047    \\
$M_{\rm 4, min}$ & \ms  &   0.047      \\
$K_4$           & \ks   &   0.52       \\
\hline
$\sum{w\ (O-C)^2}$ & day$^2$ & $ 8.5\cdot10^{-5}$  \\ 
\hline
\end{tabular}
\end{center}
\end{table}

The period analysis of \n\ was performed using all available mid-eclipse times 
found in the literature (\oc\ gateway~\footnote {\url{http://var2.astro.cz/ocgate}}, 
Paschke \& Br\'at 2006) and primarily by our long-measured series 
of mid-eclipse times. Besides those minima given in Tables~\ref{t1} and \ref{tess}, 
we used previous times of minimum obtained by 
\cite{2005IAPPP.101...38M},
\cite{2014MNRAS.442.2620Z} (their Table~4), and
\cite{2016A&A...587A..82W} (their Table~A.3).
%All precise mid-eclipse times presented in the last mention paper were
%included to our analyses.
A total of  227 precise CCD times including  110 secondary eclipses were used for the analysis.  The least-squares method was applied and resulting LITE parameters and their internal errors of the fit are given in Table~\ref{t2} (in parenthesis). 
The historical \oc\ diagram since discovery \n\ as variable star is plotted in Fig.~\ref{n745oc}. It is clearly visible that mid-eclipse times do not follow a simple linear ephemeris during past 15 years. The  current \oc\ diagrams in detail is shown in Fig.~\ref{n745ocd3}.

Assuming a coplanar orbit of the third and fourth bodies ($i_3, i_4 \simeq 90^{\circ}$) and the total mass of the eclipsing pair $M_1 + M_2 \simeq 1.45$~\ms\ \citep{2007JSARA...1....7C}, we can estimate a lower limit for the mass of the third and fourth components, $M_{\rm 3, min}$ and $M_{\rm 4, min}$.
This value, as well as the mass functions $f(m)$, and
the amplitudes of the systemic radial velocity $K_3$ and $K_4$ are also given in Table~\ref{t2}.
The probable third component may be a main-sequence star of a spectral type M3 with a bolometric magnitude of about +9.2~mag 
\citep{2013ApJS..208....9P}~\footnote{\url{http://www.pas.rochester.edu/~emamajek/EEM\_dwarf\_UBVIJHK\_colors\_Teff.txt}} produces a  detectable third light of $L_3 \simeq 6~\%$.
This is in good agreement with the solution of {\sc Tess} light curves where the third light 
of about 7~\% was found.
See Table~\ref{t3} in Section~\ref{sec:curve}.
The possible fourth body may be a brown dwarf with mass about  0.05~\ms, which is 
practically invisible in the system of two K stars.

We also tested the so-called physical delay, the direct gravitational influence of the third body on the motion of an eclipsing pair. The amplitude of the dynamical contribution of the third body, 
$A_{\rm dyn}$, is given by \cite{2013ApJ...768...33R} or \cite{2016MNRAS.455.4136B}:

$$ A_{\rm dyn} = \frac{3}{8\pi} \frac{M_3}{M_1+M_2+M_3} \frac{P_s^2}{P_3} 
             \, \left(1-e^2_3\right)^{-3/2} $$

\smallskip
\noindent
and is also listed in Table~\ref{t2}. The value of $A_{\rm dyn,3}$ is 
in the order of seconds and comparable with the uncertainty 
of individual mid-eclipse time estimation. Thus the physical delay cannot contribute 
significantly to observed changes of \oc\ values.

%%%%%%%%%%%%%%%%%%%%%%%%%%%%%%%%%%%%%%%%%%%%%%%%%%
\subsection{Magnetic activity}

A possible alternative scenario to the third body and LITE hypothesis assumes
the period modulation in binaries connected with magnetic activity of stars. 
\cite{1992ApJ...385..621A} proposed a model which explains orbital period modulations 
as a consequence of the magnetic activity changes of one of the components.
According to this model, the component, which has a magnetic activity
cycle, can show a period change of $\Delta P/P \simeq 10^{-5}$ only.

%%%%%%%%%%%%%%%%%%%%%%% 4. Light-curve solution %%%%%%%%%%%%%%%%%%%%%%%%%%
\section{Light curve solution}
\label{sec:curve}
%%%%%%%%%%%%%%%%%%%%%%%%%%%%%%%%%%%%%%%%%%%%%%%%%%%%%%%%%%%%%%%%%%%%%%%%%%

\subsection{Temperature of the primary component}
 Effective temperatures of binary components are crucial parameters of all photometric models.
In absence of direct spectroscopic results we searched available literature.
There are several different estimations of temperature found in the current catalogs which are collected in Table~\ref{temp1}. For \n\ it can be seen a difference over 2000~K. Removing both extreme values we can obtain the mean value of $4613 \pm 200$~K.

On the other hand analyzing available photometry by the more robust SED fitting using the Virtual Observatory SED Analyzer (VOSA, \cite{2008A&A...492..277B}), 
we can obtain for black-body radiation of a binary model temperatures 4300 + 4200 K. Thus we accepted the effective temperature of primary component $T_1 = 4300 \pm 100$~K, which was used in our next light curve modeling. 
\cite{2013ApJS..208....9P} give for this temperature the color index $J - H = 0.60$~mag and the spectral type K6, the {\sc Gaia} archive and the {\sc Simbad} database present the similar value  of $J - H = 0.56$~mag. 

\begin{table}
    \centering
    \caption{Temperatures of \n\ found in different sources.}
    \label{temp1}  
    \begin{tabular}{ccc}
    \hline\hline
     Source         & Temp. [K] &  Reference  \\
     \hline
    {\sc Simbad} Sp. type M2   &  3560   &    \cite{2013ApJS..208....9P}   \\
   Regression of stellar $T_{\rm eff}$ &  4503  & \cite{2019AJ....158...93B}   \\
    {\sc Tess} Input Catalog & 4533    &    \cite{2019AJ....158..138S}   \\           
    {\sc Gaia} DR2      &   4700    &       \cite{2018AA...616A...1G}   \\
    {\sc Gaia} DR3      &   4719    &       \cite{2022yCat.1355....0G}    \\
    Starhorse catalog  &   5753     &       \cite{2019AA...628A..94A}     \\
    \hline
    \end{tabular}
\end{table}

\begin{figure*}
\begin{center}
\includegraphics[width=0.32\linewidth]{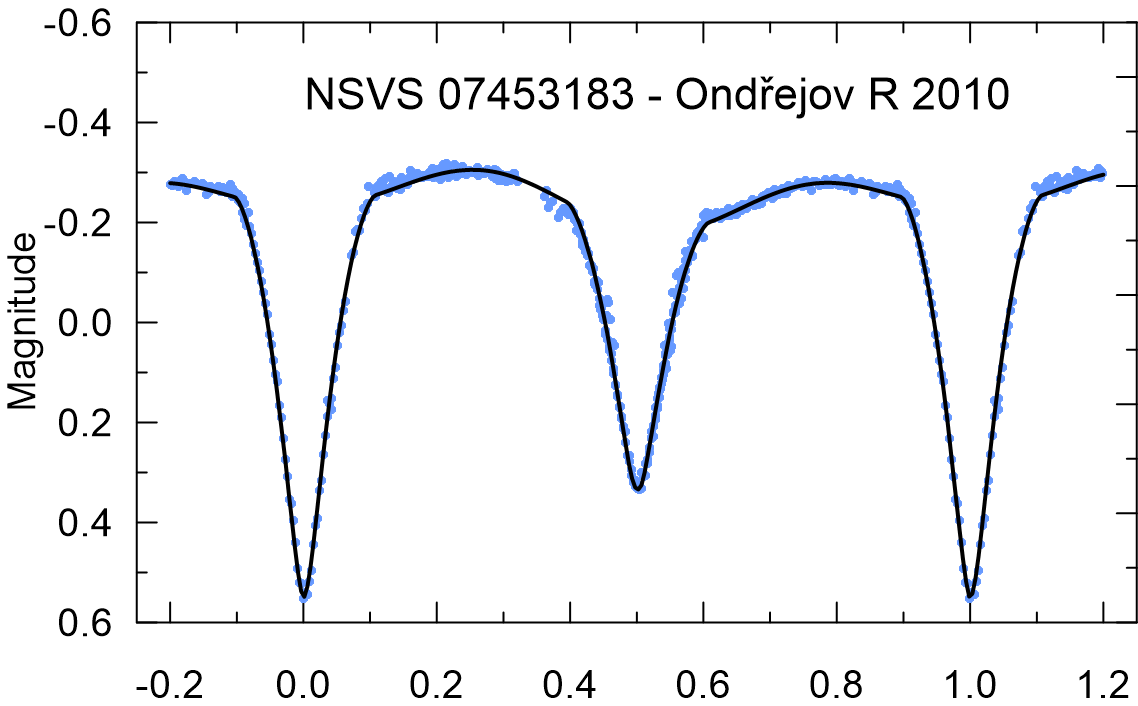}
\hspace{8 mm}
\includegraphics[width=0.32\linewidth]{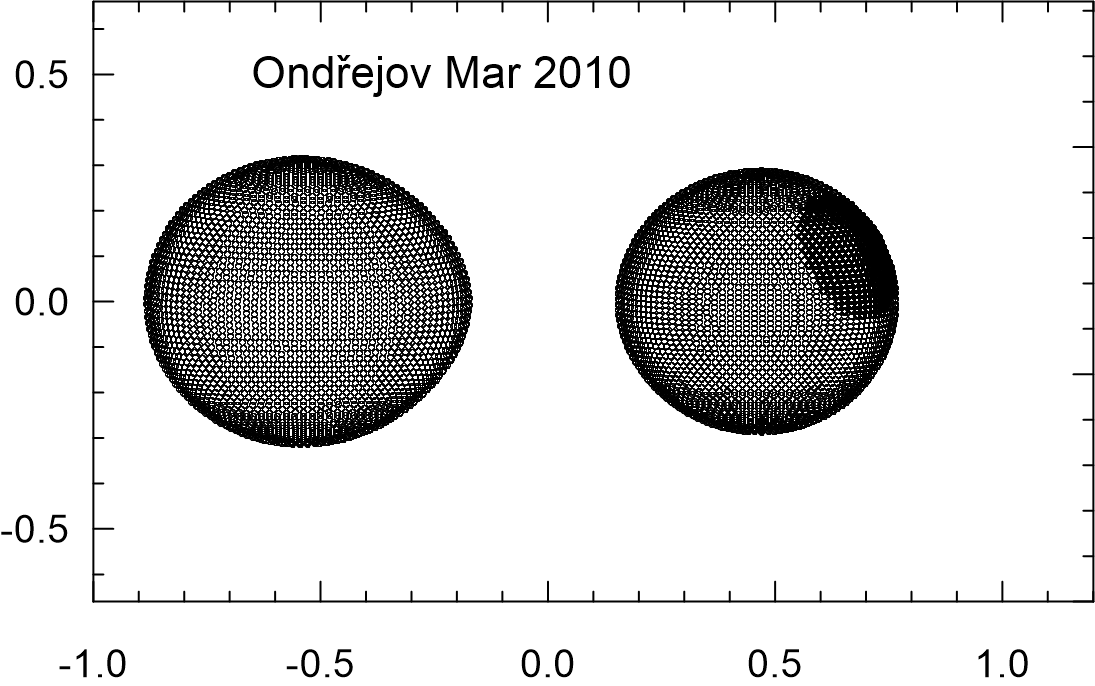}

\medskip
\includegraphics[width=0.32\linewidth]{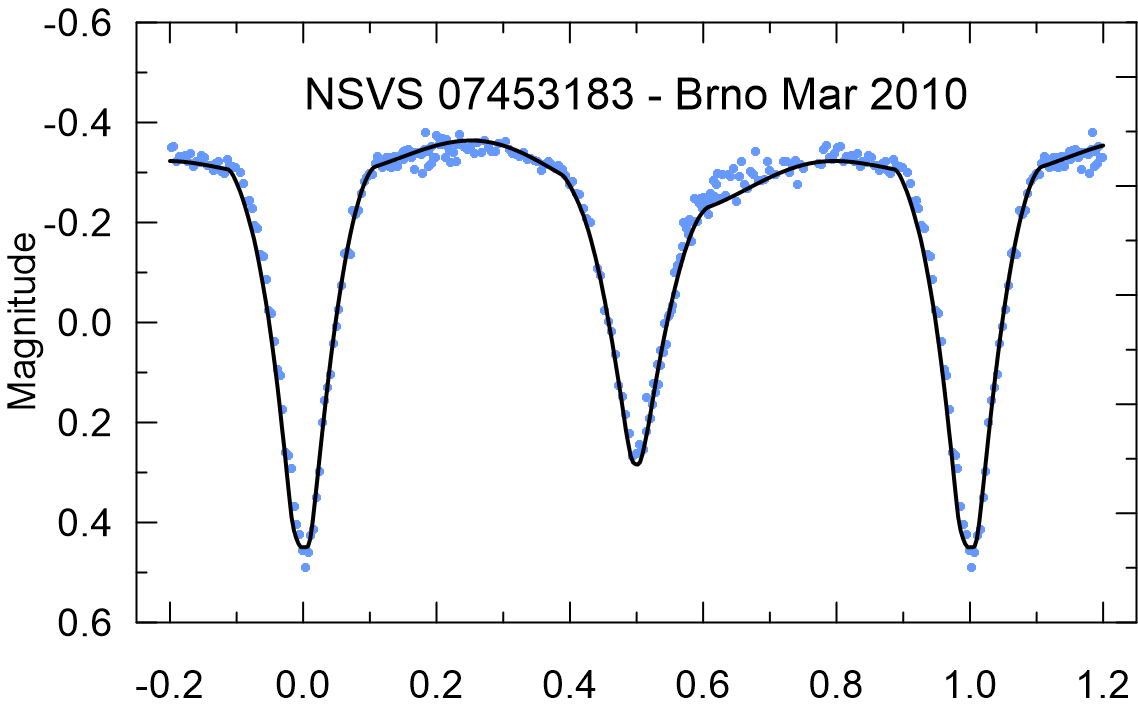}
\hspace{8 mm}
\includegraphics[width=0.32\linewidth]{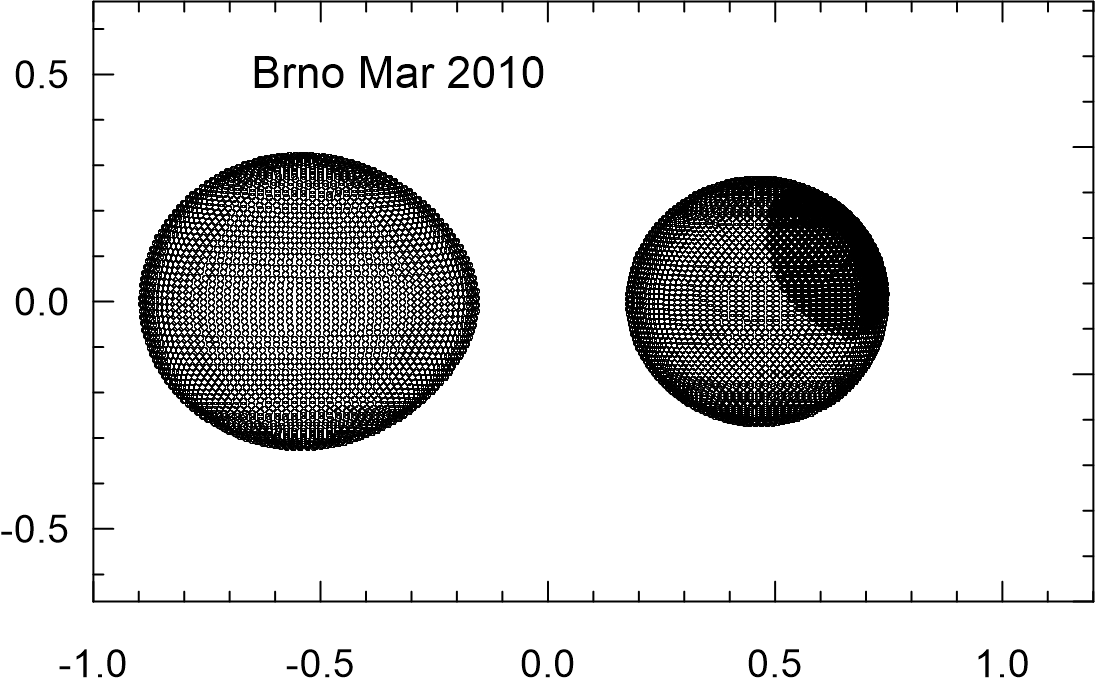}

\medskip
\includegraphics[width=0.32\linewidth]{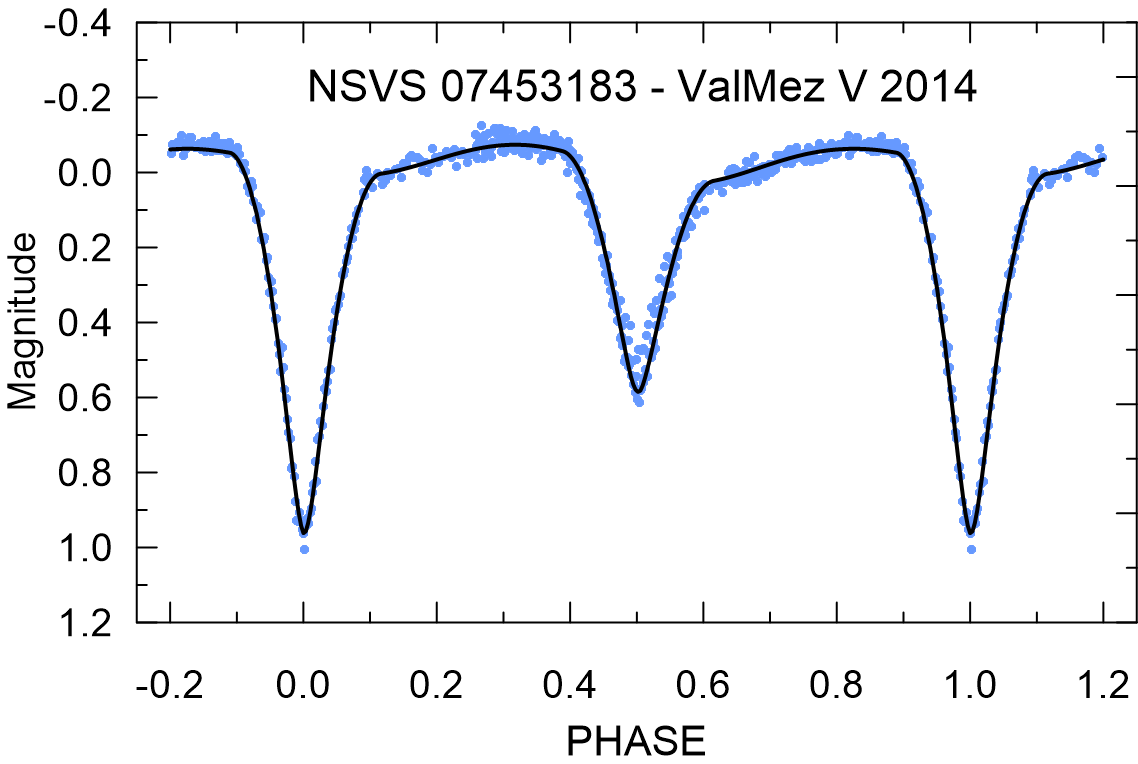}
\hspace{8 mm}
\includegraphics[width=0.32\linewidth]{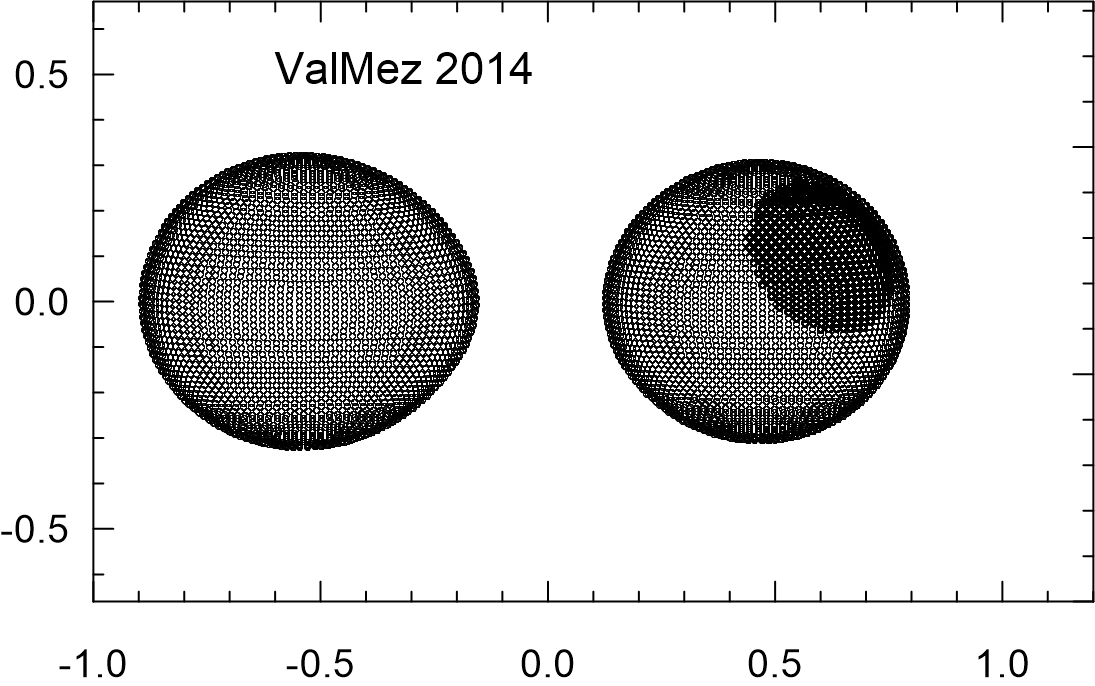}
\caption[ ] {{\bf Left:} Comparison of light curves of \n\ obtained at \ond\ (top) and Brno (middle) in 2010 and \valmez\ (bottom) in 2014 and its solution in {\sc Phoebe}. 
{\bf Right:} The geometrical configuration in phase 0.75. 
A large dark area on the surface of the primary component is visible. }
\label{comp1}
\end{center}
\end{figure*}

\begin{figure*}
\begin{center}
\includegraphics[width=0.30\linewidth]{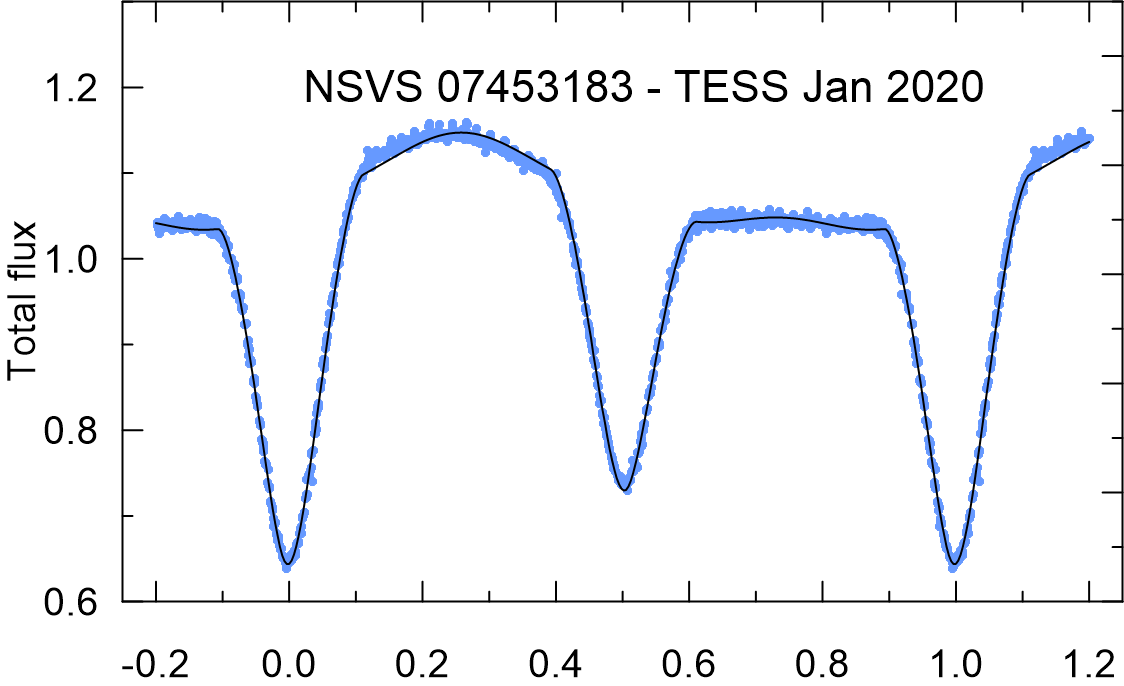}
\hspace{8 mm}
\includegraphics[width=0.30\linewidth]{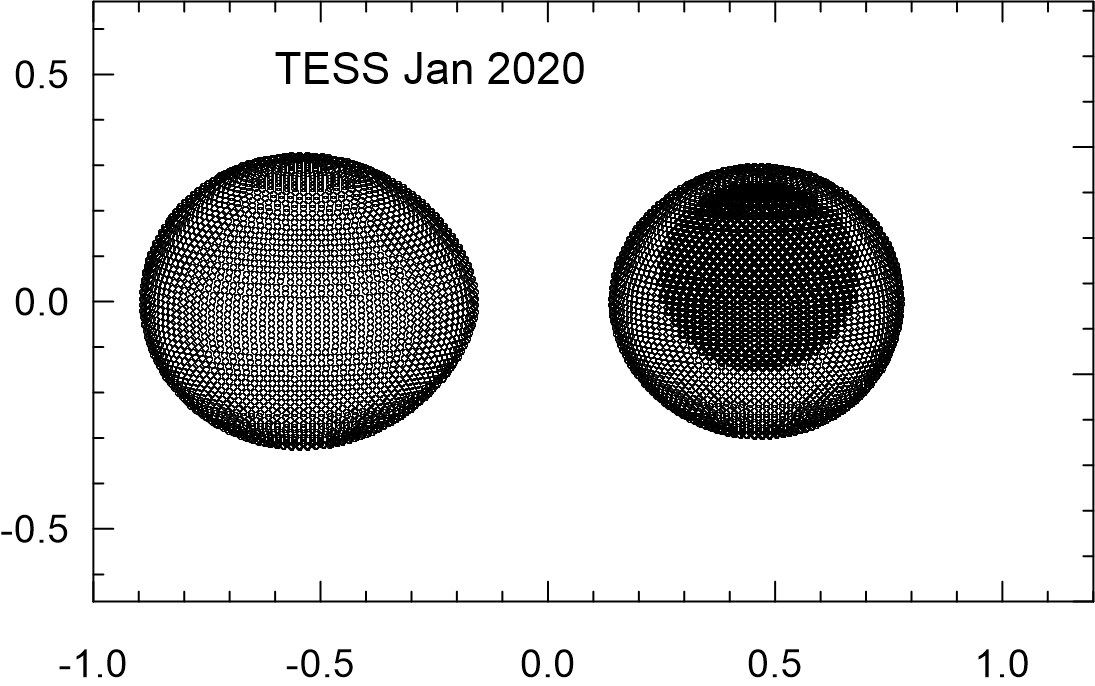}

\medskip
\includegraphics[width=0.30\linewidth]{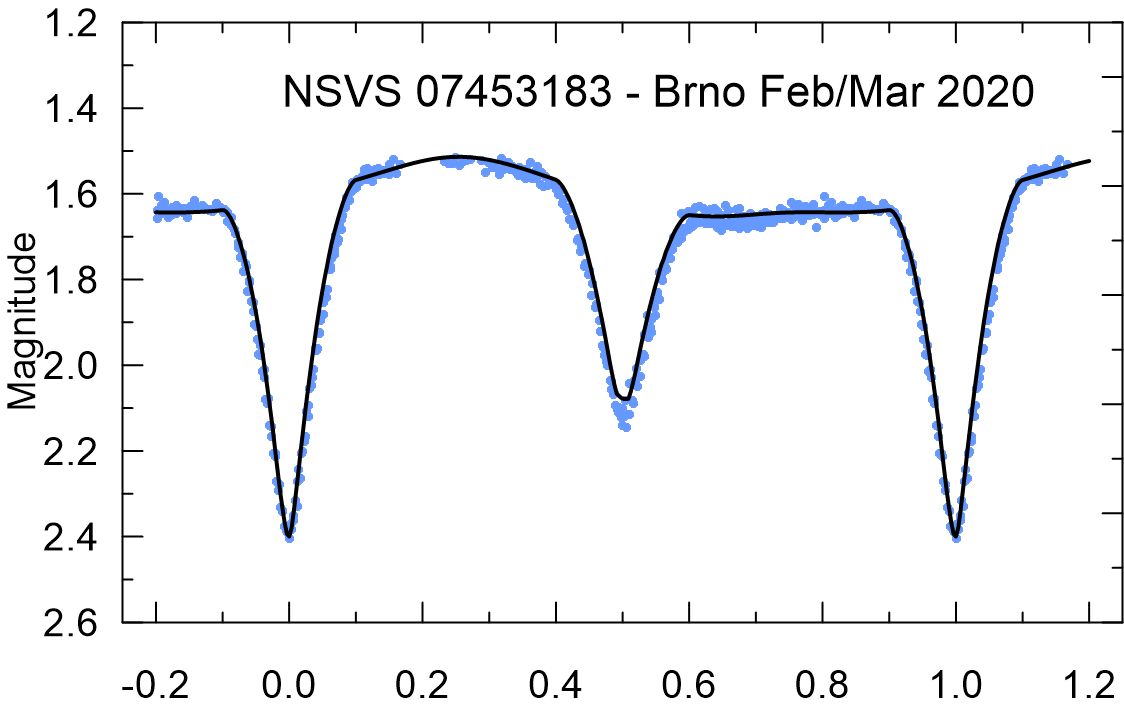}
\hspace{8 mm}
\includegraphics[width=0.30\linewidth]{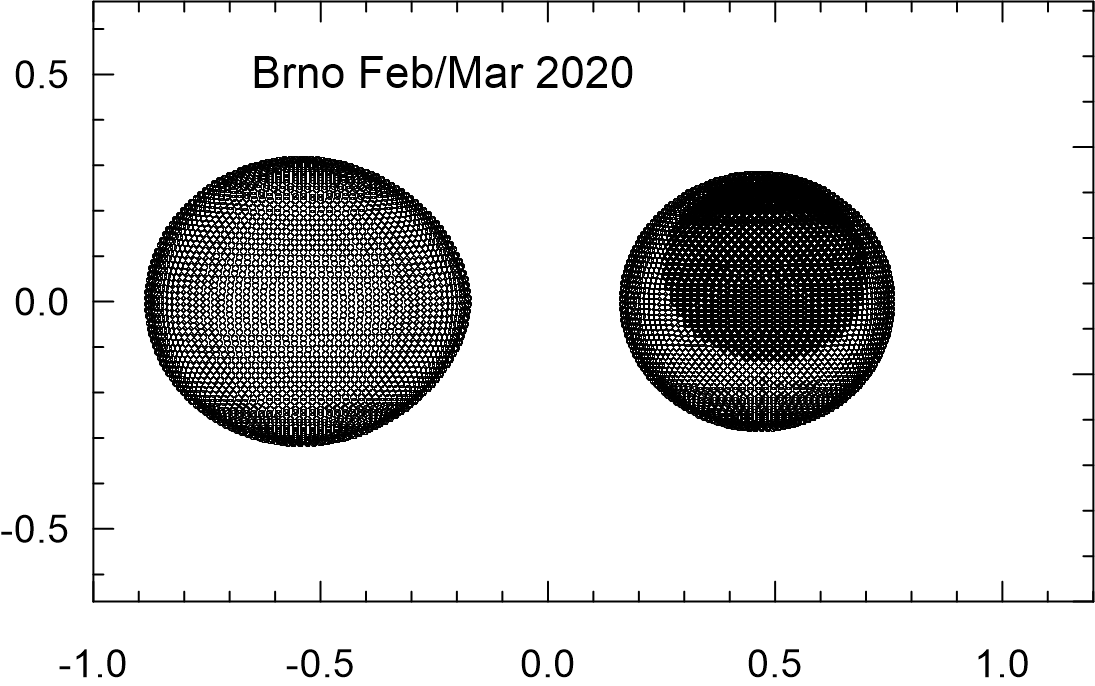}

\medskip
\includegraphics[width=0.30\linewidth]{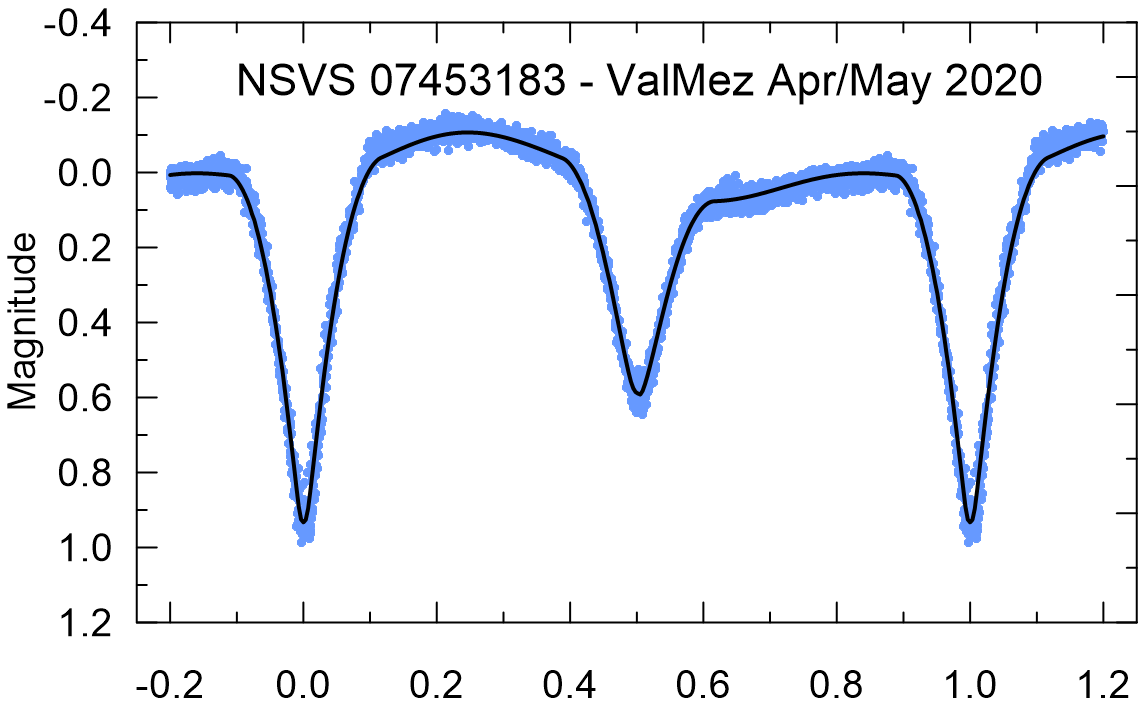}
\hspace{8 mm}
\includegraphics[width=0.30\linewidth]{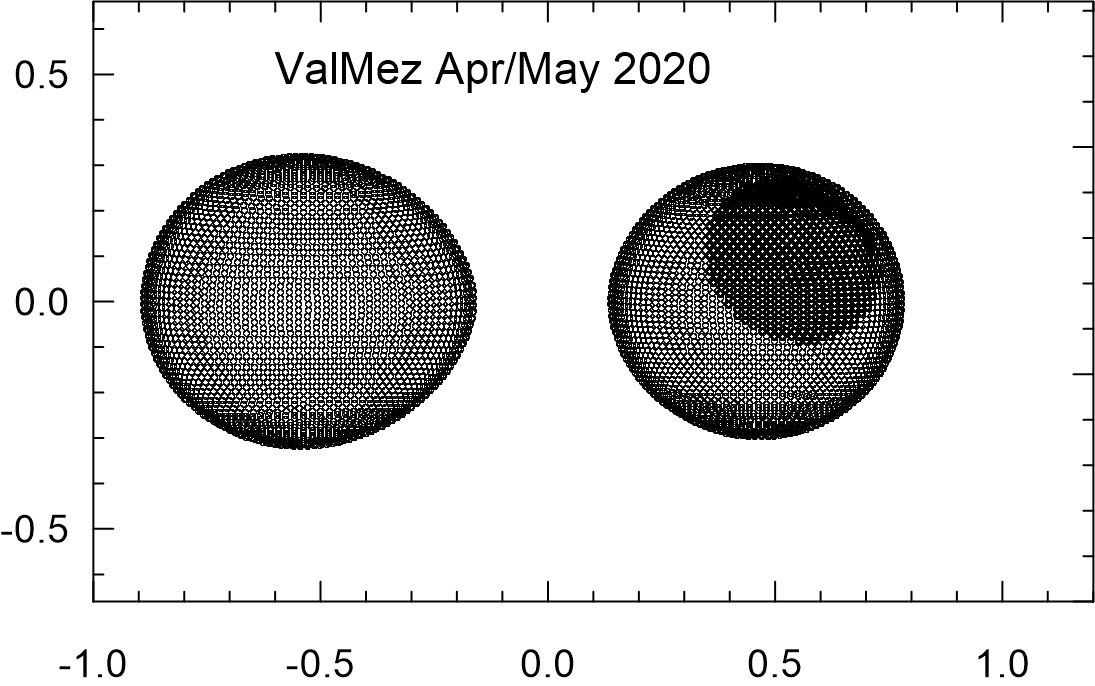}

\medskip
\includegraphics[width=0.30\linewidth]{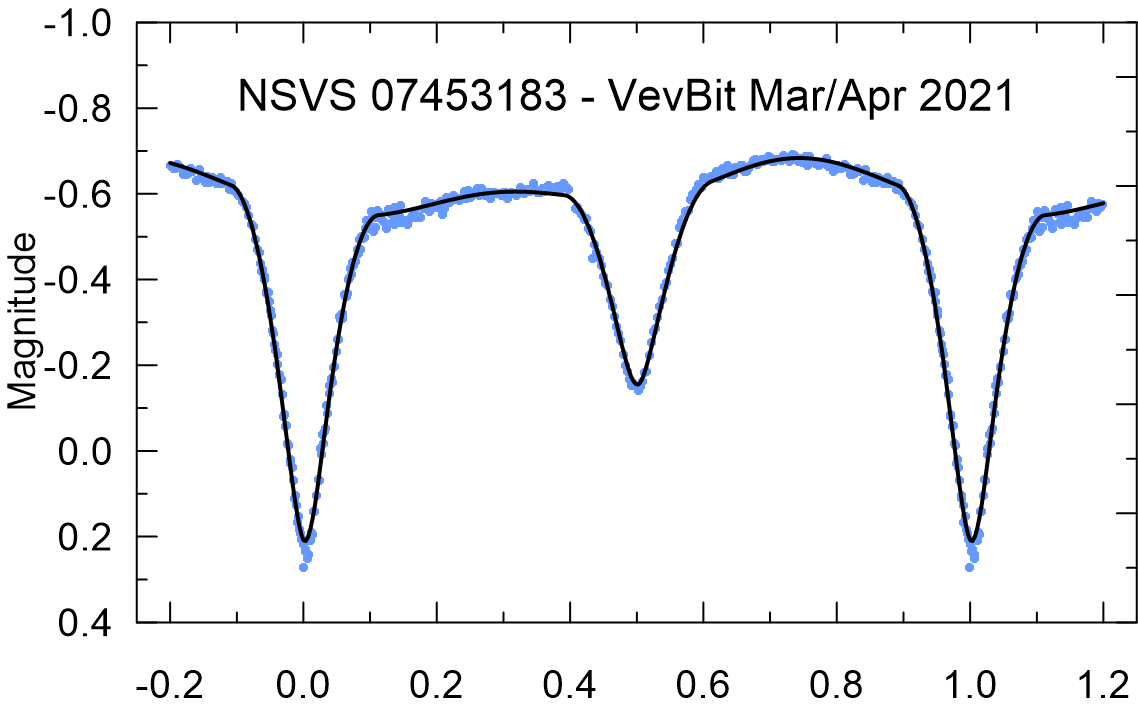}
\hspace{8 mm}
\includegraphics[width=0.30\linewidth]{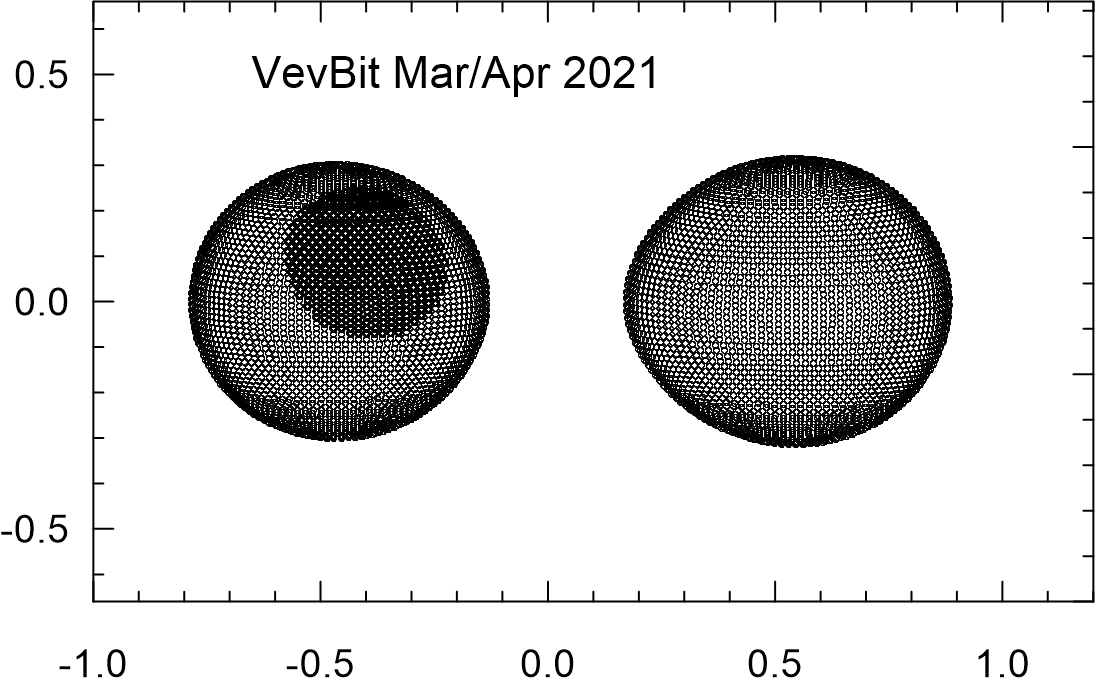}

\medskip
\includegraphics[width=0.30\linewidth]{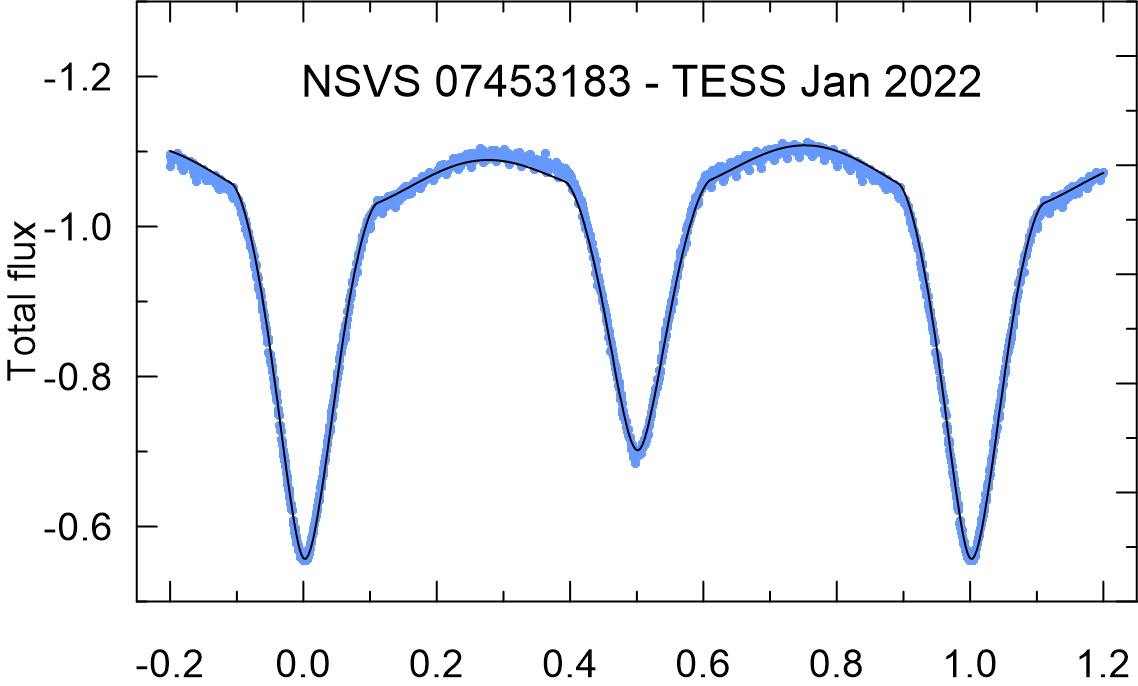}
\hspace{8 mm}
\includegraphics[width=0.30\linewidth]{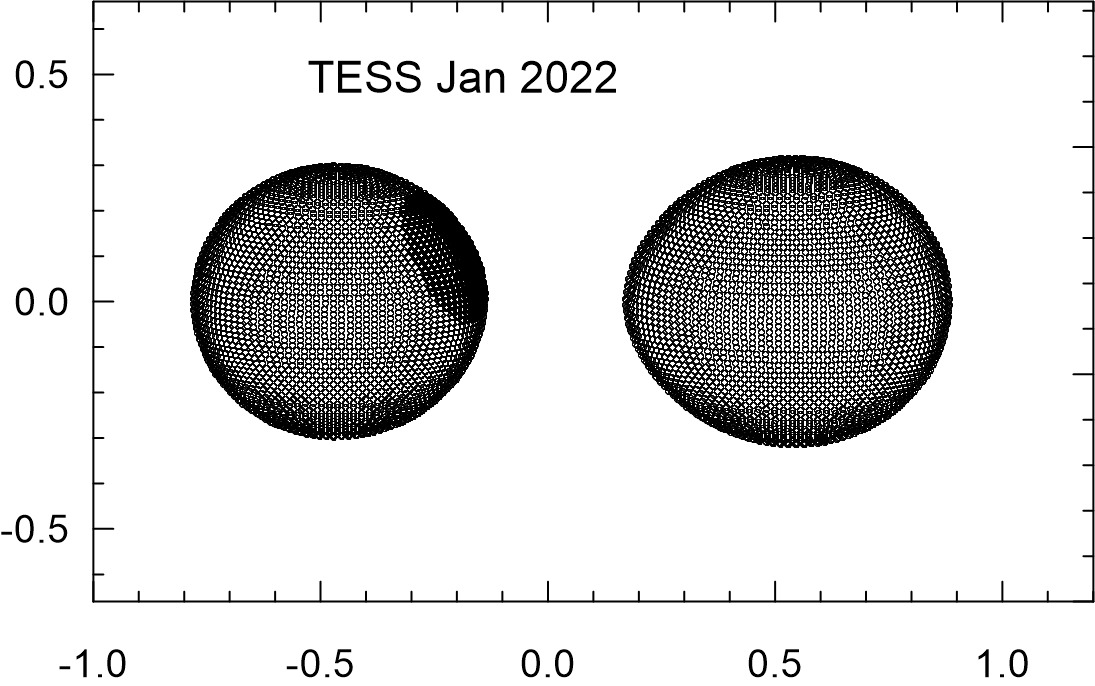}

\medskip
\includegraphics[width=0.30\linewidth]{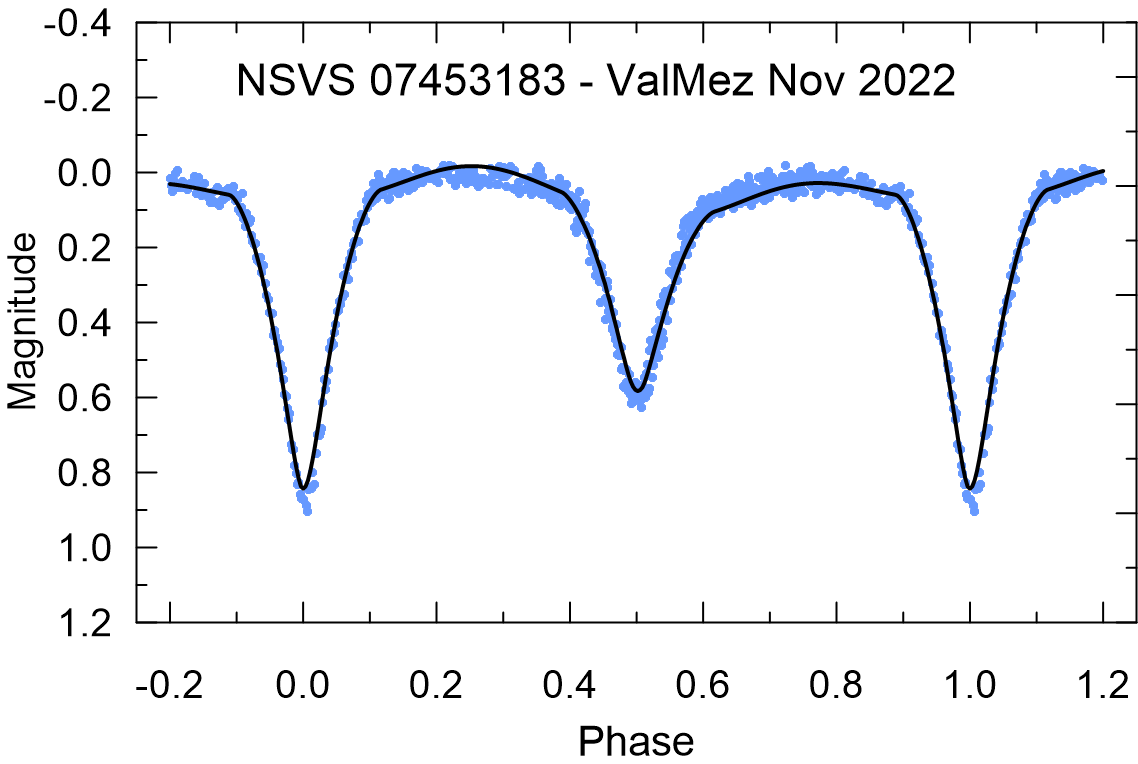}
\hspace{8 mm}
\includegraphics[width=0.30\linewidth]{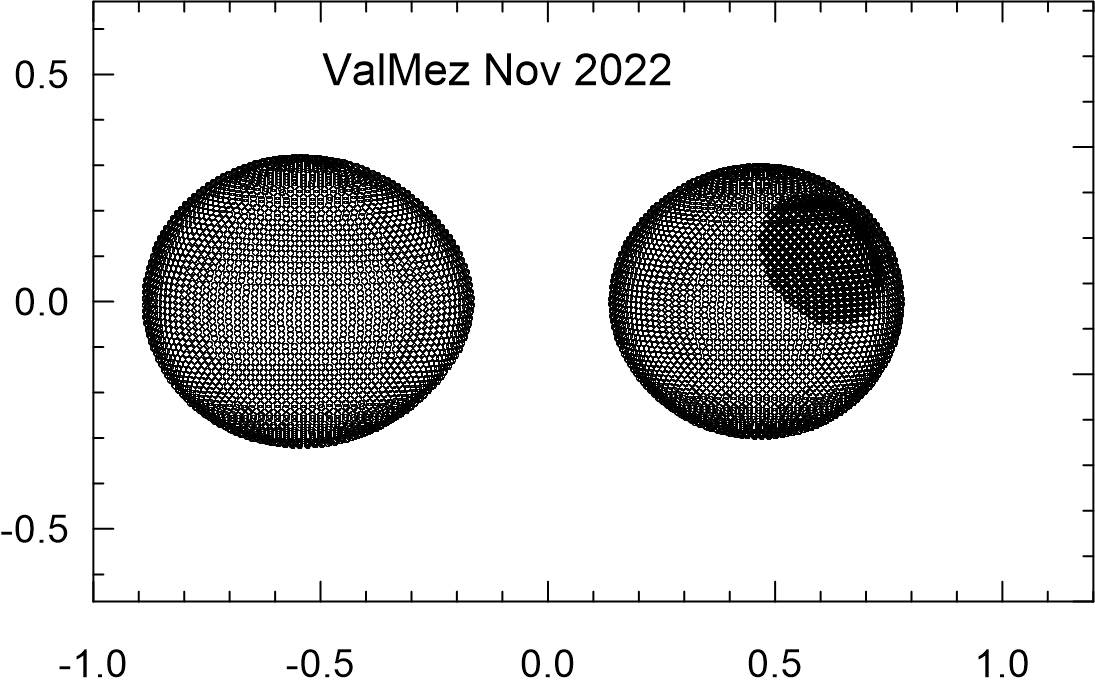}

\caption[ ]{Continuation of Fig.~\ref{comp1}. {\bf Left:} Comparison of next  six seasonal light curves of \n\ in chronological order: data from {\sc Tess} obtained in January 2020 (top), Brno and \valmez\ obtained in 2020 (middle), \vev\ in April 2021 and next {\sc Tess} photometry obtained in January 2022 and last curve from \valmez\ (bottom) and their solutions in {\sc Phoebe}. 
The different shape of the light curve is clearly visible. {\bf Right:} Corresponding geometrical models in phase 0.75 (except the 4th and 5th panel in phase 0.25). The detached configuration with the large dark region persistent on the primary component is distinguishable.}
\label{comp2}
\end{center}
\end{figure*}

\begin{table*}
\begin{center}
\caption{Comparison of photometric solutions of previous authors with our {\sc Tess} photometric elements of \n, the {\sc Phoebe} solution. }
\label{t3}
\begin{tabular}{ccccccccc}
\hline\hline
Parameter & Primary   & Secondary  &  Primary  &  Secondary 
          & Primary   & Secondary  &  Primary  &  Secondary \\
          & \multicolumn{2}{c}{\cite{2005IAPPP.101...38M}} 
          & \multicolumn{2}{c}{\cite{2014MNRAS.442.2620Z}} 
          & \multicolumn{2}{c}{TESS 2020}  
          & \multicolumn{2}{c}{TESS 2022} \\
\hline
$i$ [deg]       &  \multicolumn{2}{c}{84.8(0.1) }
                &  \multicolumn{2}{c}{84.305(0.113)}
                &  \multicolumn{2}{c}{82.2(0.5) } 
                &  \multicolumn{2}{c}{81.4(0.5) }\\
$q = M_2/M_1$   &  \multicolumn{2}{c}{1.05(0.005)}
                &  \multicolumn{2}{c}{0.601(7)}
                &  \multicolumn{2}{c}{0.861 (fixed)}
                &  \multicolumn{2}{c}{0.861 (fixed)}  \\ 
$T_{1,2}$  [K]  &  4110(10)     &  4390(10)  & 3570 (fixed) & 3438(6) 
                &   4300 (fixed) &   4075(100)  &   4300 (fixed) &   4080(100) \\
$\Omega_{1,2}$ & - & - & 4.032(44) & 3.156(20) & 3.52(12) & 3.04(15) & 3.52(12) &  3.04(15) \\
$r$(pole) &-&-& 0.2894(37)& 0.3006(34) &   0.297(5)&   0.320(7)&   0.297(4)&   0.316(6)\\
$r$(side) &-&-& 0.2953(40)& 0.3123(40) &   0.305(5)&   0.332(8)&   0.304(3)&   0.328(5)\\
$r$(point)&-&-& 0.3052(47)& 0.3675(108)&   0.323(5)&  0.377(11)&   0.322(4)&   0.369(5)\\
$r$(back) &-&-& 0.3015(44)& 0.3375(57) &   0.316(4)&  0.353(10)&   0.315(4)&   0.348(6) \\
$L_3$ [\%]& - & - & - & - & \multicolumn{2}{c} {6.4}  &  \multicolumn{2}{c}  {8.1} \\
\hline
\end{tabular}
\end{center}
\end{table*}

\subsection{ {\sc Phoebe} solution}  

As a first attempt we selected roughly 1500 points of the precise {\sc Tess} light curves 
obtained in Sectors~21 and 48, to obtain basic photometric parameters of the system.
The wavelength coverage of {\sc Tess} is about 6000 -- 10000 \AA, covering most of the R and I filters. 
These high quality light curves  were analyzed using the well-known {\sc Phoebe} code \citep{2005ApJ...628..426P, 2016ApJS..227...29P}, which is based on the Wilson-Devinney algorithm \citep{1971ApJ...166..605W} and is widely used for modeling the photometric light curves of eclipsing binaries as a standard tool. 
Because \n\ belongs to late-type binaries, we adopted the bolometric albedos and gravity darkening coefficients as $A_1 = A_2 = 0.5$ and $g_1 = g_2 = 0.32$, which corresponds to the convective envelopes (see \cite{1968ApJ...151.1123L}). Synchronous rotation for both components of the system ($F_1 = F_2 = 1$) and a circular orbit ($e = 0$) were assumed. We used the linear cosine limb-darkening law with the coefficients adopted from \cite{1993AJ....106.2096V} tables.

The adjustable parameters were the inclination $i$, the effective temperature of the secondary component $T_2$, luminosities $L_1$, $L_2$, dimensionless potentials of both components  $\Omega_1, \Omega_2$, the third light $L_3$ and the characteristics of the dark spot on the primary component (colatitude, longitude, spot radius and temperature factor).
The need to include a spot to the final solution was evident in view of the modulation of out-of-eclipse light curves.
Because the effective temperature and the radius of a spot are strongly correlated, 
we assumed that the ratio of spot/star temperatures is close to 0.9 in our analyzes.
The fine and coarse grid raster for both components were set to 40.

All additional light curves obtained \ond\ in 2010, in Brno in 2010 and 2020, in \valmez\ in 2014 and 2020, and \vev\ in 2021 were then solved independently to estimate primarily the characteristics of the dark region. 
Numerous {\sc Phoebe} runs in a detached mode using the different setup of initial 
parameters were evaluated. The results as well as the cost function value were recorded.
The final solution was accepted when subsequent iterations did not result in a decrease of the {\sc Phoebe} cost function.
 We also made some preliminary tests placing spot on both components at latitudes of 45 deg, which is the region most affected by spots in low-mass stars  \citep{2000A&A...355.1087G}.

The final light curve solution of {\sc Tess} data compared with previous results  
are given in Table~\ref{t3}, where  
%bolometric limb-darkening coefficients and 
inclination, temperatures, potentials and relative radii of both components are given.
The mass ratio was fixed to the value 0.861.
The resulting parameters of the dark region on the primary component are given in Table~\ref{t7}. 

The computed light curves based on derived parameters are compared with our measurements 
in Figure~\ref{comp1} and \ref{comp2}. The geometrical representations of \n\ in phase 0.75 are displayed on the right side of both figures.
The good visibility of dark region on the primary component close to this phase 
supports possible places of a flare origin in this area.
As one can see, the agreement between the theoretical and observed light curves 
is relatively good.

\begin{table*}
\begin{center}
\caption{Parameters of the surface structure on the primary component of \n\ obtained 
from eight different data sets.}
\label{t7}
\begin{tabular}{lccccccccc}
\hline\hline
Year                & 2010 & 2010 & 2014   & 2020  & 2020 & 2020   & 2021   & 2022 &   2022 \\ 
Filters             & VR   & VR   & V      & --    & R    & VR     & C      & --   &   R   \\ 
Observatory         & Ond  & Brno & ValMez & TESS  & Brno & ValMez & VevBit & TESS &   ValMez \\
\hline
Colatitude [deg]    & 65  &  65   &  65    &  60   & 65   & 65     & 65    & 65  &    65  \\
Longitude [deg]     & 145 &  140  &  120   &  85   & 100  & 110    & 295   & 340 &    125 \\
Spot radius [deg]   & 40  &  40   &  40    &  45   & 50   & 40     & 40    & 30  &    30  \\
%Temperature factor & 0.9 &  0.92 &  0.92  &  0.9  & 0.9  & 0.9    & 0.9   & 0.9 &  0.9 \\
Number of points    & 490 &  253  &  524   & 1150  & 431  & 2900   & 353   & 1530 &   720 \\ 
Phoebe cost function &   900 &  1200 &  1350 & 3150 & 5650 & 2900  & 640   & 4150 &  1150 \\    
\hline
\end{tabular}
\end{center}
Observatory:  Ond = \ond, ValMez = \valmez, VevBit = \vev
\end{table*}

%%%%%%%%%%%%%%%%%%%%%%%%%%%%%%%%%%%%%%%%%%%%%%%%
\section{Flare activity}
\label{sec:flare}
%%%%%%%%%%%%%%%%%%%%%%%%%%%%%%%%%%%%%%%%%%%%%%%%

\begin{figure*}
\begin{center}
\includegraphics[width=0.6\linewidth]{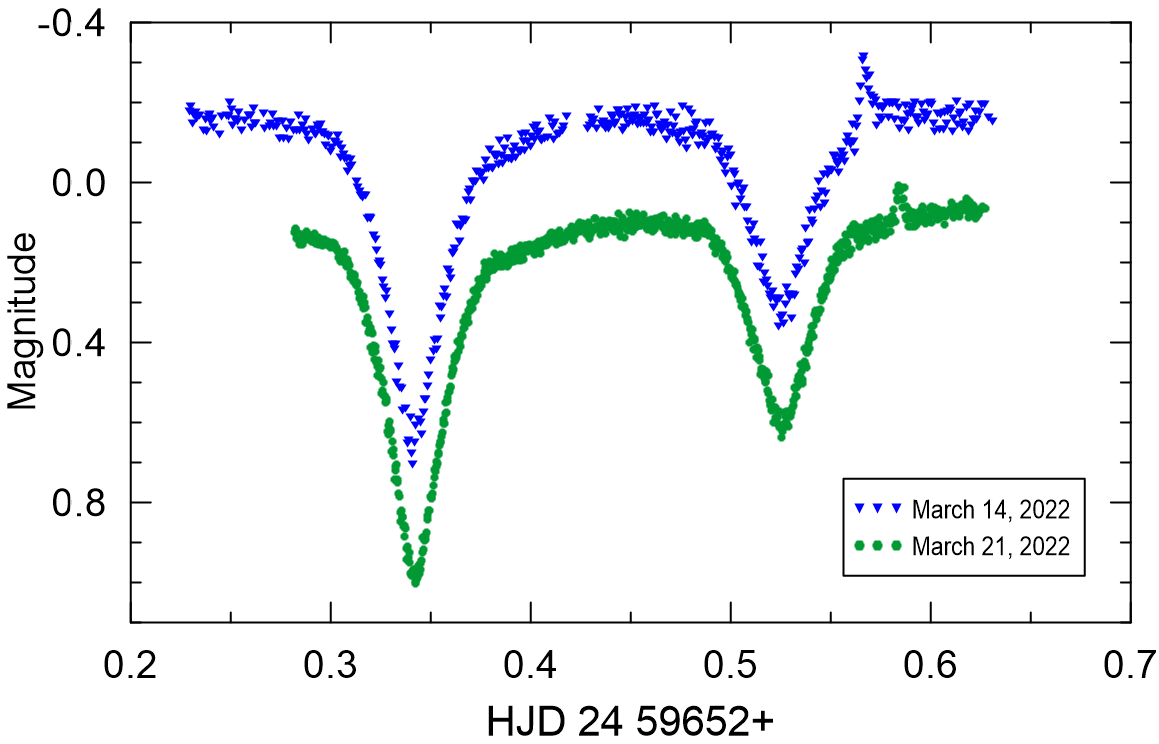}
\caption[ ]{Typical light curves of \n\ obtained in clear filter at \valmez\ observatory on 2022, March 14 (blue triangles) and March 21 (green circles), both showing primary and secondary eclipses followed by a small flare on the similar phase. The amplitude of flares are 0.19 and 0.09~mag, respectively. The light curve obtained on JD 24~59659 was shifted and phased with the previous one for clarity.}
\label{N745erup}
\end{center}
\end{figure*}

Stellar flares are sudden events in stellar atmospheres, 
which eject a hot plasma into surroundings and release a quantum
of accumulated magnetic energy. The total released energy
in the flares ranges usually from $10^{24} - 10^{27}$ J \citep{1989SoPh..121..299P}.
We can remind, that the significant eruption activity is well known in several 
low-mass eclipsing binaries: 
CM~Dra \citep{1977ApJ...218..444L, 1997IBVS.4462....1K, 2007IBVS.5789....1N, 2009Ap.....52..512K}, 
YY~Gem \citep{1990A&A...232...83D}, 
V405~And \citep{2009A&A...504.1021V},  
DV~Psc \citep{2010NewA...15..362Z}, 
NSVS~6550671 \citep{2010MNRAS.406.2559D}, 
CU~Cnc \citep{2012MNRAS.423.3646Q}, and 
GJ~3236 \citep{2017MNRAS.466.2542S}.

Eruption activity of \n\ was first reported by \cite{2014MNRAS.442.2620Z}.
They observed only one flare event appeared around phase 0.39 on HJD 24~55904.3064. 
The flare duration was found to be 116~min (phase 0.33-0.55) and the time required 
for the flare to peak was about 32~min. The amplitudes of the flare were 0.1 mag in V, 
0.076~mag in R and 0.05~mag in I band. 
They also noted, that flare might have been connected to the primary component 
due to visibility during this phase.
\cite{2014MNRAS.442.2620Z} monitored the object on Dec 8, 2011, and during 26.6 hours only 
one eruption has been found. The flare frequency was thus derived 0.0376 eruption per hour.
Moreover, the hydrogen emission lines are strong indicators of the chromospheric activity
(see Fig.~\ref{keckspec})

In our photometric campaign, \n\ was measured every clear night at \valmez\ observatory 
from Jan 22, 2019 to May 26, 2022. 
Typical light curves obtained in clear filter are plotted in Fig.~\ref{N745erup}.
Altogether 107 observing nights and 594.3 hours of CCD photometry were analyzed. 
We employed the same flare detection criterion as in our previous
study of the active eclipsing binary GJ~3236 \citep{2017MNRAS.466.2542S}. 
During this time interval only 6 new eruptions were counted and a new flare frequency 
of 0.00168 eruption per hour was derived. This is 4~times less than was found 
in the previous study of \cite{2014MNRAS.442.2620Z}. 

Detailed information about observed flares are summarized in Table~\ref{eruptab}.  
Surprisingly, four eruptions were observed during relatively short time span in March~2022.
Having such a small sample, the energies of individual flares were not calculated. 

\begin{table}
\begin{center}
\caption{Parameters of flares observed at \valmez\ observatory. }
\label{eruptab}
\begin{tabular}{ccccccc}
\hline\hline
No. & Max JD Hel. & Event & Duration &  Amplitude &  Filter    \\
    & --2400000   &        & [min]    &  [mag]     &            \\
\hline
1 & 59161.56148 &   Max &   4.3 &    0.13 &    V         \\
  & 59161.56223 &   Max &   3.1 &    0.06 &    R         \\
2 & 59309.29322 &   Max &  13.5 &    0.06 &  Clear       \\
3 & 59640.45301 &   Rise &  2.5 &         &              \\
  &             &   Max &   7.5 &    0.09 &  Clear       \\       
4 & 59641.26040 &   Rise & 17.0 &         &              \\
  & 59641.27220 &   Max &  48.0 &    0.07 &  Clear       \\
5 & 59652.59490 &   Rise &  5.2 &                        \\
  & 59652.59853 &   Max &  14.6 &    0.19 &  Clear       \\
6 & 59659.58468 &   Rise &  4.0 &                        \\
  & 59659.58745 &   Max &  10.5 &    0.09 &  Clear       \\
\hline
\end{tabular}
\end{center}
\end{table}

%%%%%%%%%%%%%%%%%%%%%%%% 5. Discussion  %%%%%%%%%%%%%%%%%%%%%%%%
\section{Discussion}
\label{sec:discus}
%%%%%%%%%%%%%%%%%%%%%%%%%%%%%%%%%%%%%%%%%%%%%%%%%%%%%%%%%%%%%%%%%%%%%%%%

The physical parameters of the binary suggest that the times of the tidal synchronization and orbital circularization are very short. These times can be derived as 
\citep{2001icbs.book.....H}

\begin{equation}
t_{\rm sync} \simeq 10^4 \, \left[ \frac{1+q}{2q} \right]^2 P^4,  
\end{equation}

\begin{equation}
t_{\rm circ} \simeq 10^6 \, q^{-1} \left[ \frac{1+q}{2} \right]^{5/3} P^{16/3}, 
\end{equation}

\noindent
where times are in years, $q = M_2/M_1$ is the mass ratio and $P$ is the orbital period in days. For \n\ these times are
$t_{\rm sync} \simeq 10^3$~yr and
$t_{\rm circ} \simeq 10^5$~yr.
These values are very small compared to typical ages of low-mass binaries 
and justify our assumptions about the orbital circularization and synchronicity parameters.

\begin{figure}
    \centering
     \includegraphics[width=0.95\linewidth]{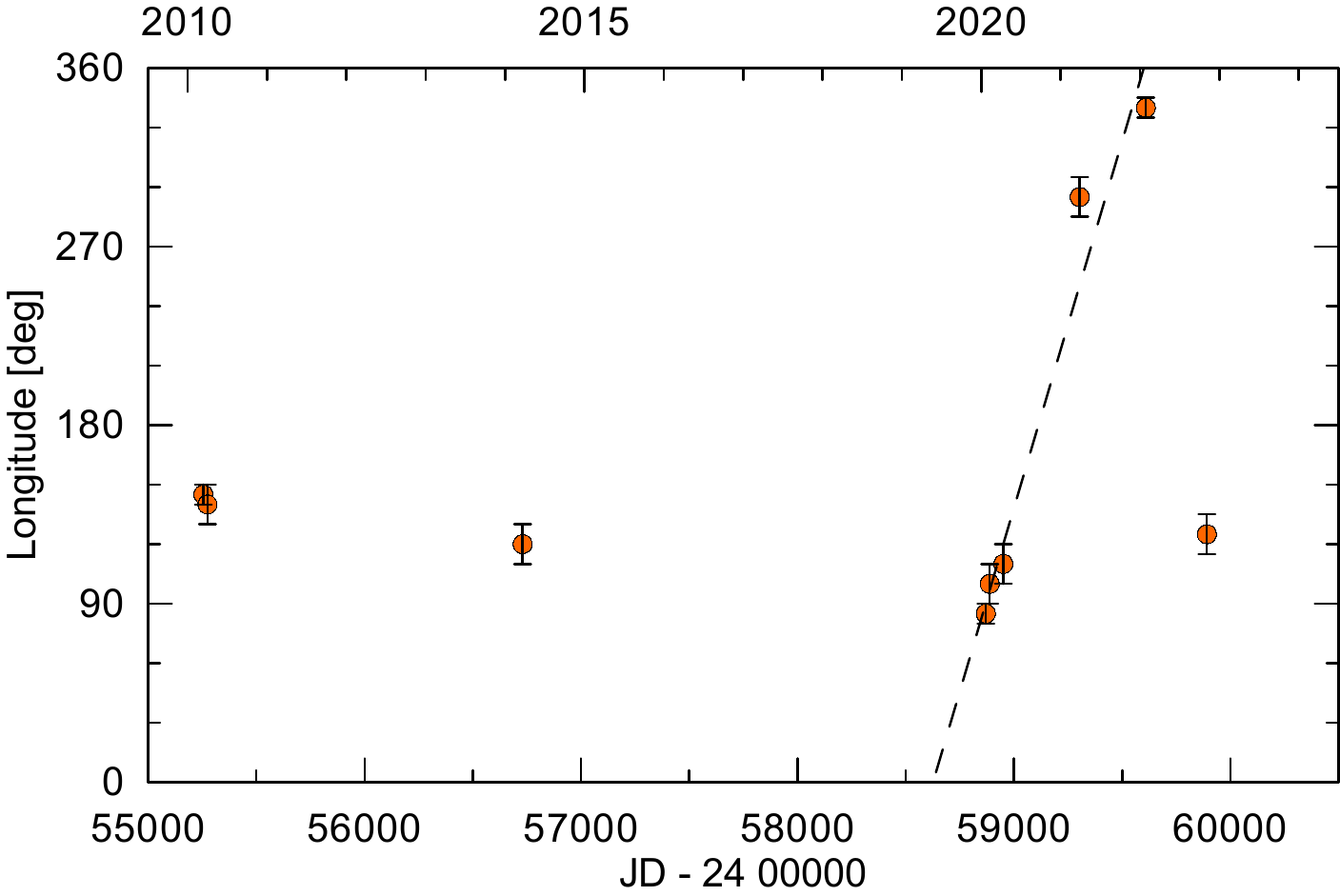}
    \caption{The changes in longitude of the dark structure on the surface of the primary component during several observational epochs. Mean errors in longitude are indicated. 
    The steep dashed line denotes migration of the dark structure during 2020-2022 with the rate about 10 deg/month}.
    \label{745spot}
\end{figure}

The seasonal light curves of \n\ show slowly evolving and persistent starspot structures that also strongly affect the estimation of precise \oc\ timings.
The apparent period changes are better visible on the run of secondary eclipses and are probably caused by a moving dark structure on the surface of the primary component, where a different part of this region is visible. 
Position of the dark region derived from available light curves are collected in Table~\ref{t7}.
 Fig.~\ref{745spot} shows changes in the spot longitude during 12 years including two epochs of {\sc Tess} photometry.  Although the resulting colatitude and the radius of the dark structure remain practically the same, the value of its longitude has been changing over the last few years.
 This is also well visible on the current \oc\ diagram (Fig.~\ref{n745ocd3}), where secondary minima show a significant shift up to 0.002~days from the predicted values.
 Due to sparse coverage of this diagram on Fig.~\ref{745spot} we also cannot exclude a rapid change in the longitude of this cold structure found on light curves of \n\ during last years 2020-2022.
 The average migration speed of about 10 deg/month was derived which is close to the value recently derived by \cite{2022NewA...9701879S} for another low-mass eclipsing binary V608~Cam.
 
The sinusoidal changes of the dark region longitude in time were described by \cite{2014MNRAS.442.2620Z} in another low-mass eclipsing binary system NSVS~2502726. They derived variation with the period $5.9 \pm 0.2$ yr with an amplitude of about 33 deg. \cite{2013ApJ...774...81T} studied eclipse timing variations in contact binary systems from the {\sc Kepler} archive. They showed that the anticorrelated behavior, the amplitude 
of the \oc\ delays, and the overall random walk-like behavior can be explained by the presence  of a starspot that is continuously visible around the orbit and slowly changes its longitude on timescales of weeks to months. The quasi-periods of $\sim 50-200$ days were observed in the \oc\ curves.
Later, \cite{2015MNRAS.448..429B} reported rotational motions of spots on the surfaces of components in very short-period, near-contact, and contact binary systems selected from 414 {\sc Kepler} light curves. They found that in $\sim 34$ \% of the systems, the spot rotation was retrograde as viewed in the frame rotating with the orbital motion, while $\sim 13$ \% showed significant prograde spot rotation. The remaining systems showed either little spot rotation or erratic behavior, or sometimes include intervals of both types of behavior.

%%%%%%%%%%%%%%%%%%%%%%%% 6. Summary  %%%%%%%%%%%%%%%%%%%%%%%%
\section{Conclusions}
\label{sec:concl}
%%%%%%%%%%%%%%%%%%%%%%%%%%%%%%%%%%%%%%%%%%%%%%%%

A study of late-type and low-mass binaries provides us with important information about 
the most common stars in our Galaxy. The next rare system with a large dark region on 
the surface of the primary component was studied. 
New ground based and long-term photometric observations of \n\ supplemented by the 
precise {\sc Tess} photometry were used to determine photometric parameters of the 
components and to investigate its asymmetric light curve. 
New mid-eclipse times were determined and up-to date \oc\ diagram was constructed.
The third body orbiting the eclipsing pair announced in our previous study \cite{2016A&A...587A..82W} 
was confirmed and more precise LITE elements were obtained.
Moreover, our results indicate that \n\ is a next quadruple system in an ((1+1)+1)+1 
hierarchy. The detached eclipsing binary is orbiting by a third body with the relatively 
short period of 425~days only. The fourth body, probably a brown dwarf, has an orbital 
period of about 12~years.   

The \oc\ diagram assembled on Fig.~\ref{n745oc} is one of the most detailed description 
of the period changes of a low-mass eclipsing binary. Except LITE due to two additional bodies it shows also systematically shifted secondary eclipses caused by surface activity.
The longitudinal migration of dark structures on the surface 
of components is relatively frequent and was studied recently by \cite{2021RMxAA..57..351Y} 
or \cite{2022NewA...9701879S}.
We can conclude that during years 2020-2022 the motion of the dark structure 
on the surface of the primary component was prograde and also relatively rapid 
with the average migration speed of about 10 deg/month.

The third light computed from the light-curve solution (7\%) is in good agreement with the value derived from the minimal mass of the third body. 
The detected very low eruption activity of \n\ is remarkable in comparison with 
similar known low-mass stars.  

Long-term systematic monitoring of this object can bring more information about nature 
of period changes and surface activities. New high-dispersion and high-S/N spectroscopic observations are needed to obtain the radial-velocity curves and to derive accurate masses for this interesting multiple system. It is also a challenge for theoreticians to clarify 
the origin and dynamical stability of this hierarchical quadruple system with 
a brown dwarf as the most distant orbiting body.

\section*{Acknowledgments}

Useful suggestions and recommendations by an anonymous referee helped us to improve 
the clarity and correctness of the paper and are greatly appreciated.
The research of MW and PZ was partially supported by the project 
{\sc Cooperatio - Physics} of the Charles University in Prague.
HK and KH were supported by the project RVO: 67985815.
The authors would also like to thank 
Lenka Kotkov\'a, \ond\ observatory, 
Marek Chrastina, Masaryk University Brno,
Jan Vra\v{s}til, Tereza Je\v{r}\'abkov\'a, and Ond\v{r}ej Chrenko, 
all former students of the Charles University in Prague, for their important contribution 
to photometric observations during past decades. 

This paper includes data collected by the {\sc Tess} mission.
Funding for the {\sc Tess} mission is provided by the NASA Science Mission directorate. 
Some of the data presented in this paper were obtained from the Mikulski Archive for Space
Telescopes (MAST).
This work has made use of data from the European Space Agency (ESA) mission {\sc GAIA}~\footnote {{\sc Gaia}, \url{https://www.cosmos.esa.int/gaia}}, processed by the {\sc Gaia} Data Processing and Analysis Consortium 
(DPAC)~\footnote{{\sc Dpac}, \url{https://www. cosmos.esa.int/web/gaia/dpac/consortium}}. 
Funding for the DPAC has been provided by national institutions, in particular 
the institutions participating in the {\sc Gaia} Multilateral Agreement. 
This research has made use also of the Keck Observatory Archive (KOA), 
which is operated by the W.M. Keck Observatory and the NASA Exoplanet Science Institute (NExScI), under contract with the National Aeronautics and Space Administration.
 This publication makes use of {\sc Vosa}, developed under the Spanish Virtual Observatory~\footnote{{\sc Vosa}, \url{https://svo.cab.inta-csic.es}} project funded by MCIN/AEI/10.13039/501100011033/ through grant PID2020-112949GB-I00.
{\sc Vosa} has been partially updated by using funding from the European Union's Horizon 2020 Research and Innovation Programme, under Grant Agreement No. 776403 (EXOPLANETS-A).
The following internet-based resources were used in research for this paper:
the SIMBAD database operated at CDS, Strasbourg, France,
the NASA's Astrophysics Data System Bibliographic Services, 
and the O-C Gateway of the Czech Astronomical Society.
%This investigation is part of an ongoing collaboration 
%between professional astronomers and the Czech Astronomical Society, 
%Variable Star and Exoplanet Section.

%%%%%%%%%%%%%%%%%%%%%%%%%%%%%%%%%%%%%%%%%%%%%%%%%%
\section*{Data Availability}

Some of the data were derived from sources in the public domain, 
and the respective URLs are provided as footnotes. 
The other data are available on reasonable request to the authors.



\bibliographystyle{mnras}
\bibliography{example} % if your bibtex file is called example.bib

%\begin{thebibliography}{99}
%\end{thebibliography}

%%%%%%%%%%%%%%%%%%%%%%  End of paper %%%%%%%%%%%%%%%%%%%%%%%%%%%%%%%%%%%%%%%%%%%%%%%%%%%%%%%%  
%%%%%%%%%%%%%%%%%%%%%%%%%%%%%%%%%%%%%%%%%%%%%%%%%%%%%%%%%%%%%%%%%%%%%%%%%%%%%%%%%%%%%%%%%%%%
% Don't change these lines
\bsp	% typesetting comment
\label{lastpage}
\end{document}